# Electronic Properties of Single Prussian Blue Analog Nanocrystals Determined by Conductive-AFM.

Hugo Therssen[1], Laure Catala[2], Sandra Mazérat[2], Talal Mallah[2], Dominique Vuillaume[1], Thierry Mélin[1], Stéphane Lenfant[1]

1. Institut d'Electronique de Microélectronique et de Nanotechnologie (IEMN), CNRS, Univ. Lille, 59652 Villeneuve d'Ascq, France.
2. Institut de Chimie Moléculaire et des Matériaux d'Orsay (ICMMO), CNRS, Université Paris-Saclay, 91400 Orsay Cedex, France.

**Abstract**

We report a study of the electron transport (ET) properties at the nanoscale (conductive-AFM noted C-AFM thereafter) of individual Prussian Blue Analog (PBA) cubic nanocrystals (NCs) of $CsCo^{III}Fe^{II}$, with size between 15 and 50 nm deposited on HOPG. We demonstrate that these PBA NCs feature an almost size independent electron injection barriers of $0.41 \pm 0.02$ eV and $0.27 \pm 0.03$ eV at the $CsCo^{III}Fe^{II}$/HOPG and $CsCo^{III}Fe^{II}$/C-AFM tip, respectively, and an intrinsic electron conductivity evolving from largely dispersed between $\sim 5\times10^{-4}$ and $2\times10^{-2}$ S/cm without a clear correlation with the nanocrystal size. The conductivity values measured on individual nanocrystals are higher by up to 5 decades than those reported on PBA films.

**Introduction**

Prussian Blue Analog (PBA) nanocrystals are nano-objects at the frontier between molecules and bulk materials, and have molecular properties that can be used for different applications[1,2] such as gas storage[3,4,5], materials for energy issues,[6,7,8,9] magnetic properties for information storage,[10] electrochemical and biosensors,[11] catalysis,[12] environmental purification,[13,14] or biomedical applications.[15] The structure of PBA consists of a cyanide-bridged bimetallic face-centered cubic (fcc) arrangement obtained from the reaction of a hexametallocyanate (negatively charged) and a hexa-aquometallate of transition metal ions (Fe in the pristine Prussian Blue) upon the substitution of water molecules by the nitrogen atoms of the metallocyanate molecules.[16] The neutrality of the network is ensured either by the presence of metallocyanate vacancies and/or by the insertion of positively charged alkali ions in the tetrahedral sites of the fcc structure. A wide variety of PBAs has been synthesized using different transition metal ions such as Cr, Mn, Co.[17,18] At the nanoscale, nanocrystals were prepared in water allowing the stabilization of nanocrystals with well-defined size and shape,[19] that led using a seed mediated approach to the design of core-multishell heterostructures.[20] The same approach is used here to prepare nanocrystals of increased size.

Here, we present an electrical study of the electronic transport (ET) through isolated PBA nanocrystals composed of $Cs\{Co^{III}[Fe^{II}(CN)_6]\}$ noted CsCoFe thereafter. Only few results exist in the literature, which describe ET properties of PBA. In particular the electron injection energy barriers at the contact interface and the intrinsic conductivity of PBA NCs have not been investigated so far. Previous experiments consisted in either ET experiments conducted at the microscale on film or powders,[21,22,23] or results obtained in our group at the nanoscale[24] demonstrating a weak attenuation of the electron transport through several (up to 3) CsCoFe NCs (15 nm in size) connected in series between two electrodes, i.e. a low electron transfer decay factor $\beta$=0.11 nm$^{-1}$, with the current I $\propto$ e$^{-\beta d}$, d the distance between the electrodes (15<d<45 nm). The current decay in these multi-NCs devices was ascribed to a multi-step coherent tunneling between adjacent NCs with a strong energy coupling (0.1-0.25 eV) between the adjacent CsCoFe NC.[24] Here, we report ET in individual CsCoFe NCs with various sizes in the same range (nominally: 15 nm, 30 nm and 50 nm). We prepared and deposited the CsCoFe NCs on highly ordered pyrolytic graphite (HOPG) surfaces and we characterized the electronic transport through individual NCs, i.e in HOPG/single PBA/C-AFM tip devices as a function of the PBA NC size. We found that the current increases with the NC size. To analyze the experimental current-voltage curves, we used two models. We first considered a quantum transport model as in our previous study of ET in one to three 15 nm NC in series.[24] In this single energy level (SEL) model,[25,26] the electrons are transferred between the two electrodes through a single molecular orbital (MO). Since the NC size is large compared to atomic- and molecular-scale (for which the SEL model was initially developed), we have also used a model suitable for nanoscale devices, i.e. a nano-Schottky diode as already used for C-AFM-tip/semiconductor nanocrystals, quantum dots and nanostructures interfaces.[27,28,29] More specifically, we used a simple analytical model that consists in a double Schottky barrier (DSB) model[30] (one Schottky diode at each interface) in series with the intrinsic resistance of the PBA NCs. The main observed features are the following:

i) The SEL model gives an energy level of the MO involved in the ET at about 0.3-0.5 eV below the Fermi energy of the electrodes, independent on the NC size, in agreement with the values in Ref. [[24]].

ii) The increase of the current with the NC size is not well explained by this model.

iii) With the DSB model, we found a Schottky barrier height (SBH) for hole injection of 0.41 ± 0.02 eV at the CsCoFe/HOPG interface and 0.27 ± 0.03 eV and CsCoFe/C-AFM PtIr tip, independent of the NC size. This finding is in contrast with size-dependent SBH observed in various non-ideal Schottky diodes,[31,32,33] or diodes with nanometer-scale contacts.[34,35] and nano-diodes made of semiconducting nanocrystals or nanodots.[27,28,29]

iv) the intrinsic electron conductivity of the CsCoFe NCs is largely dispersed between ~ $5 \times 10^{-4}$ S/cm and $2 \times 10^{-2}$ S/cm without correlation with the NC size.

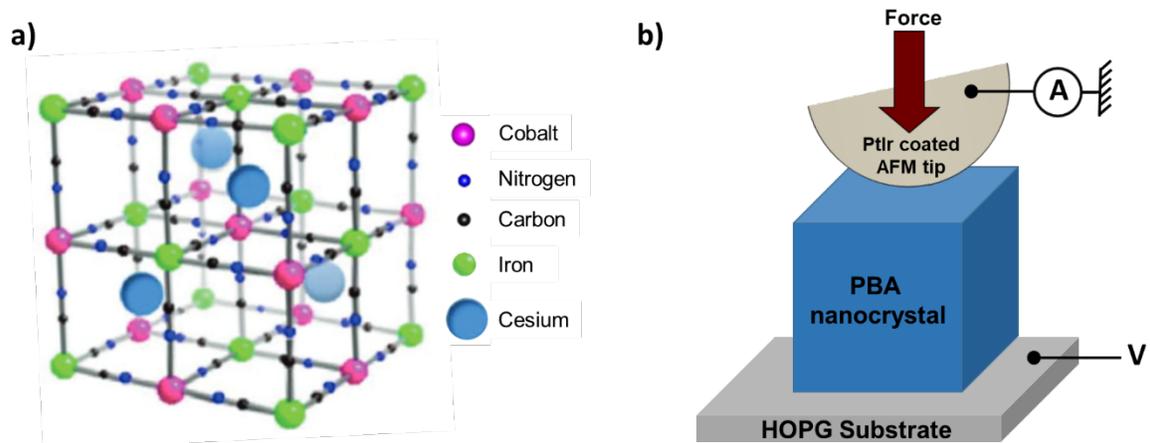

*Figure 1. (a) Scheme of the unit cell of the fcc structure of CsCoFe PBA NCs (adapted from [24]). The cell parameter is close to 10 Å. (b) Scheme of the conducting AFM experiments, in which an individual PBA NC with a characteristic size around 50 nm deposited on a freshly cleaved HOPG substrate is addressed, and statistics of I(V) characteristics are recorded using a given AFM tip force (see text). The AFM PtIr coated tip is grounded, the voltage V is applied on the HOPG substrate.*

**Experimental section**

*Sample preparation*. The synthesis of the 15 nm CsCoFe nanocrystals has been reported elsewhere.[24,36] The larger size objects were obtained by the same approach used for the similar CsNiCr NCs (see details in the Supporting Information, section 1).[37] The CsCoFe NCs are then deposited on a HOPG substrate by immersing a freshly cleaved HOPG surface in the colloidal suspension for 20 seconds. This step is followed by a thorough rinsing of the sample with deionized water and then with methanol, and finally a drying step under vacuum for several hours. The samples appeared stable over time in ambient air and at room temperature, but have nevertheless been kept in nitrogen environment for storage purposes.

*C-AFM measurements* were realized in an air-conditioned laboratory ($T_{amb}$ = 22.5 °C, relative humidity of 35-40 %), with a Dimension Icon microscope (Brüker, US) equipped with a Nanoscope V controller and a PF-TUNA module with sensitivity factors comprised between 20 pA/V to 100 nA/V. Conductive (PtIr metal plated) tips have been used, both for topography imaging and I(V) measurements (SCM-PIT-V2 from Brüker with an apex radius ~ 25 nm, a spring constant k = 2.8 N/m, and a resonance frequency of 75 kHz). Contact-mode AFM was not used for sample imaging since it leads to the displacement of NCs outside the scanning area, even using a low scanning force (a few nN), due to the weak interaction force between the PBA NCs and the substrate. We instead imaged the sample topography in tapping mode, and addressed individual NCs electrically by switching from tapping to contact mode with the tip almost in point mode above a given NC. More precisely, the measurement procedure is conducted as follows: (i) the sample topography is acquired in tapping mode (TM-AFM); (ii) an isolated PBA NC is selected; (iii) the TM-AFM scan is progressively focused on the NC top (final scan size below 10 x 10 nm²) ; (iv) the AFM oscillation driving amplitude is suppressed to stop the cantilever oscillation; (v) the AFM tip is then gently brought into contact with the NC top, while monitoring the applied force as from the cantilever static deflection ; (vi) I(V) characteristics are recorded using a typical tip force of 15 nN (experimental data consist of a set of about 100-400 I(V) curves successively obtained by applying voltage ramps between -1 V to +1 V in the forward and then backward direction). All the I-V curves were acquired with the C-AFM tip positioned on the top surface of the PBA NC; (vii) after the I(V) data acquisition, the sample topography is imaged back in tapping mode to check for the integrity of the individual NC. For a given NC size, 2 to 3 NCs were measured on the same sample using the above-described I(V) measurement scheme.

**Results**

*Topography by AFM after the NC layer deposition*

Typical TM-AFM and scanning electron micrographs (SEM) images of the CsCoFe NCs for the different sizes and freshly deposited on HOPG substrates from a colloidal solution are presented in Figure 2. TM-AFM images confirmed the presence of NCs on the HOPG substrates with a sub-monolayer coverage. The surface coverage, however, differs from sample to sample (due to the simplistic deposition method). The samples with the 15 nm

NCs have a surface coverage of ca. ~220 NCs/µm², with a histogram/grain size analysis shown in Figure 2b given an average NC size of 15.8 ± 4.9 nm. The surface with the 30 nm NCs presents a lower surface coverage of around 57 NCs/µm², with an average NC height of 26.9 ± 7.2 nm and for the 50 nm NCs, we obtained ca. ~110 NCs/µm² and an average NC height of and 42.4-± 18.1 nm. We notice a clear increase of the dispersion of the NCs size with the increase of the nominal size of the NC, this is due to the mutli step synthetic method to prepare nanocrystals with increased size (see SI for details). All samples (with nominal NC size of 15 nm, 30 nm and 50 nm) on HOPG substrates enabled to access individual NCs, as seen from the AFM images (and cross-sections) shown in figures 2d and 2e. Accessing individual NCs in a controlled way is a prerequisite to conduct ET transport measurements. In Figure 2d, an isolated nanocrystal of 15 nm, 30 nm and 50 nm of CsCoFe are presented with the corresponding height profile. We find the nominal height for each nanocrystal with, however, a larger lateral size due to the convolution effect with the radius of curvature of the AFM tip (estimated by the supplier around 25 nm).

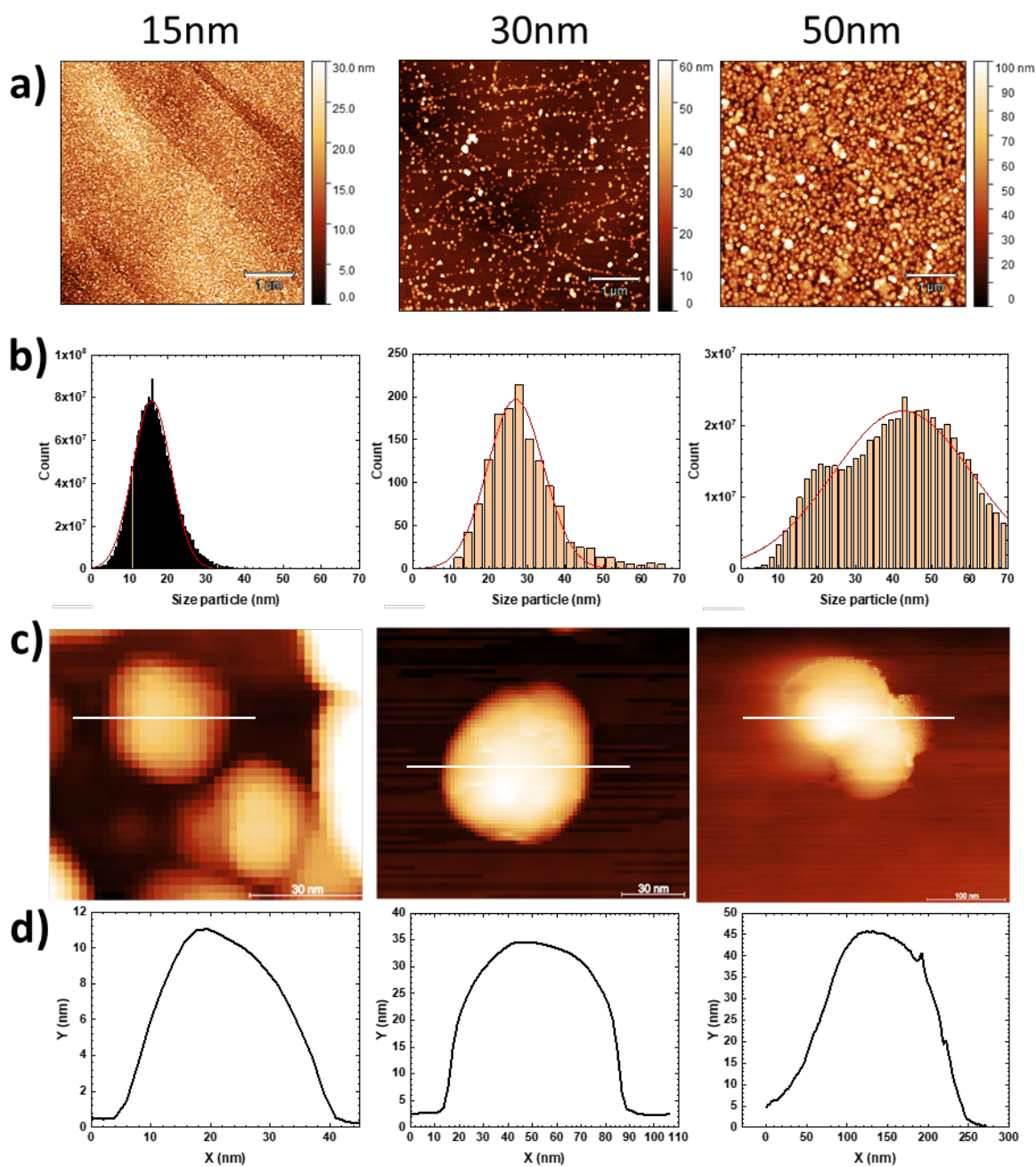

***Figure 2.*** *(a) 5 x 5 µm topographic TM-AFM images for the three sizes of CsCoFe NCs (15 nm, 30 nm and 50 nm) deposited on a HOPG substrate; (b) Corresponding height histograms (function Statistical Functions / Distribution of heights in Gwyddion v2.60 software) (15 nm and 50 nm NCs). Histograms for the 30 nm NCs have been obtained using a grain recognition analysis (function Data Process / Grains / Mark by Edge Detection with a Laplacian fixed at*

*55% in Gwyddion v2.60 software) due to the lower NC surface coverage. The average heights and the standard deviations have been obtained from the fit of the histogram with a Gaussian distribution (red lines). (c) TM-AFM images of single NCs with (d) the corresponding cross sections.*

*Electronic transport properties by C-AFM*

Figure 3 illustrates the evolution of current-voltage I(V) characteristics of NCs as a function of their size. For each size of CsCoFe NC, the current-voltage I(V) 2D histograms (Figure 3, middle) reveal low dispersed current values. The current histograms at fixed voltage (here +/- 0.4 V) are fitted with a log-normal distribution given a log-mean current (log-$\bar{I}$) and log-standard deviation log-$\sigma$. We note a low log-$\sigma$ (lower than 0.3, see Table 1) indicating a reproducible and controlled acquisition of the I(V) characteristics with the C-AFM. By comparison, the I(V) 2D histogram realized on the same HOPG substrate (as a reference) presents higher current levels (e.g. log-$\bar{I}$ > -8.5, $\bar{I}$>3x10$^{-9}$ A, at 0.4 V) (see Figure S2 in SI), thus the I(V) characteristics shown in Figure 3 correspond to the ET properties through the PBA NCs. The I(V) curves on NCs appear almost symmetric, with low ratios (inferior to 3 measured at +/- 1 V). The same measurements were done on other NCs deposited on the same HOPG samples, the corresponding datasets are shown in section 3 in the Supporting Information, Fig. S3). Figure 4 summarizes the evolution of the mean current $\bar{I}$ (at +/- 1V for all the measured samples).

|  |  | 15 nm NC | 30 nm NC | 50 nm NC |
|---|---|---|---|---|
| **- 0.4 V** | **log-$\bar{I}$** | -10.73 | -10.35 | -9.31 |
|  | **$\bar{I}$ (A)** | 1.86 x 10$^{-11}$ | 4.47 x 10$^{-11}$ | 4.90 x 10$^{-10}$ |
|  | **log-$\sigma$** | 0.18 | 0.23 | 0.26 |
| **+ 0.4 V** | **log-$\bar{I}$** | -10.76 | -10.76 | -9.51 |
|  | **$\bar{I}$ (A)** | 1.74 x 10$^{-11}$ | 1.74 x 10$^{-11}$ | 3.09 x 10$^{-10}$ |
|  | **log-$\sigma$** | 0.25 | 0.15 | 0.25 |

*Table 1: Parameters of the log-normal fits of the current distributions at -0.4V and +0.4V (Figures 3, left and right respectively): log-mean current (log-Ī), the corresponding mean current Ī, and the log-standard deviation (log-σ).*

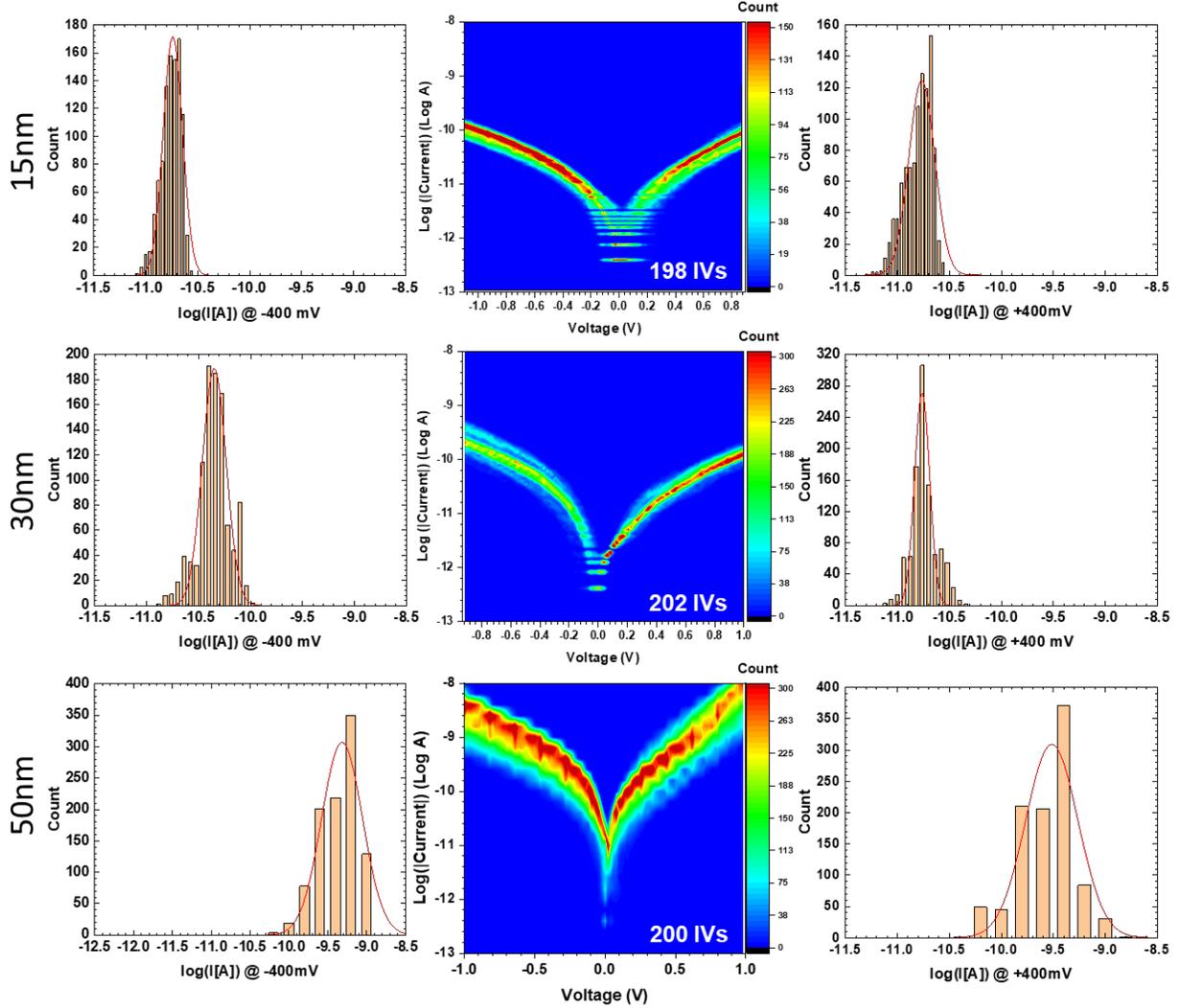

*Figure 3. C-AFM 2D current-voltage I(V) histograms obtained with around 200 IVs traces (middle) and corresponding 1D histograms at -0.4 V (left) and +0.4 V (right) for three NCs of 15 nm, 30 nm and 50 nm deposited on bare clean HOPG (dataset #1). The average current at +/- 0.4 V (log-Ī) and the log-standard deviation (log-σ) were obtained on each 1D histogram from the fit with a Gaussian distribution (red lines), see fitted values in Table 1.*

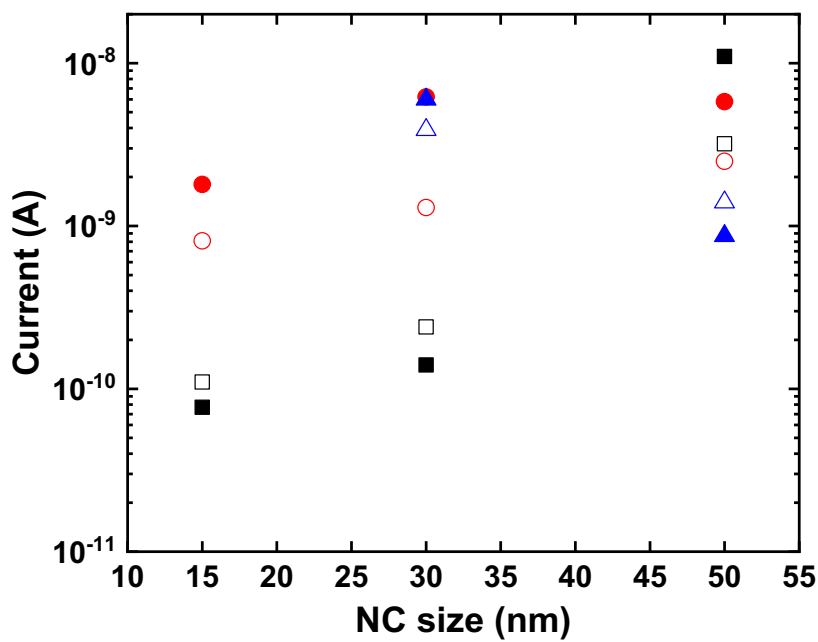

*Figure 4*. Mean current at 1V (filled symbol) and -1V (open symbol) for all the measured data set: for data shown in Fig. 3 (black square symbols), dataset #2 in Fig. S3 (red circle symbols), in blue dataset #3 in Fig. S3 (blue triangle symbols).

**Discussion**

Figures 3 and 4 show that the current increases with the NCs size increasing from 15 nm to 50 nm nominal size, however, with a large dispersion from sample-to-sample. We started analyzing the I(V) datasets with the SEL model.[25,26] In this model, the ET occurs through a single MO (here the HOMO according to our previous results)[24] located at an energy $\varepsilon_0$ below the Fermi energy of the electrodes. The MO is coupled (hybridized) to the electrodes, and these hybridizations are characterized by two coupling energies, $\Gamma_1$ and $\Gamma_2$ (see details in the Supporting Information). Fitting all the average I(V) of all the datasets shown Fig. 3 and Fig. S3, we found $\varepsilon_0$ values between ∼ 0.3 and 0.5 eV, independently of the NC size (Figs. S7 and S8 in the SI). These values are consistent with those obtained for the one-to-three 15 nm NC in series and they were ascribed to the HOMO of the CsCoFe NC ($Fe^{2+}$-$t_{2g}$ d orbital). However, for the single PBA devices, we observed an increase of the current with the NC size in contrast to the current decay previously observed in the multi-NCs devices.[24] This latter behavior was explained by a strong coupling between the adjacent NCs and an ET dominated by hopping between them,[24] a mechanism not relevant here. In the case of single PBA NCs, with the SEL model the current increase is accounted for by an increase (a factor ∼ 7) of the coupling energies $\Gamma_1$ and $\Gamma_2$ with the NC size (Fig. S9 in the Supporting Information). Since the same electrodes (HOPG and PtIr tip) were used, the same NC/electrode interfaces were formed, it is difficult to understand why the interface hybridization would depend on the NC size. We note that the values of $\Gamma_1$ and $\Gamma_2$ in the present study are in the same range (here ∼ 0.2-0.8 meV) as in our previous work (∼ 0.5-1.5 meV)[24] supporting the fact that the same interactions with the HOPG and C-AFM tip are involved in the case of a single PBA NC of various sizes and in the case of interconnected multi-NCs as in Ref.[24] Moreover, the SEL model cannot be applied to the whole applied voltage range between -1V and 1V (see details on the validity of the SEL model in the Supporting Information), we considered another model.

Materials of the PBA family are considered as large band gap semiconductors. Several works (on bulk and thin films of various PBA materials) have reported a band gap of around 2 eV[38,39] and a temperature activated behavior of electrical conductivity (in their partly oxidized state, the vacuum dried materials are insulators).[22,40,41,42,43] Given the size of the NCs, between the molecular scale and microscopic devices, the concept of nano Schottky is relevant, which takes into account non-ideal SB due to the low dimensionality of the metal/semiconductor interface (Refs.[27,28,29,35], see also a mini review in [44] and references therein). In these nano-Schottky diodes, the SBH values differ from those in their macroscopic counterparts, and ideality factors deviating far from unity (e.g. up to 2.5)[28] were observed. In particular, we found that the I(V) experimental datasets are well described by a double SB model[30] recently proposed to explain ET in several nanoscale devices (graphene[45], $MoS_2$[46]). The double SB model consists of two back-to-back Schottky diodes at the interfaces in series with the intrinsic resistance of the NC (Fig. 5a): one diode accounts for the HOPG/CsCoFe interface ($\phi_{B1}$) and the other one ($\phi_{B2}$) for the interface at the CoCsFe/PtIr C-AFM tip. From the fit of I(V) curve, this analytical model allows extracting the two SB energy simultaneously albeit the two SBs are dissimilar (i.e. different energy barriers, ideality factors and contact areas),[30] and the intrinsic resistance of the NC. In this model, the current is always limited by the saturation current of the reverse-biased diode, the other diode being in the forward regime. Here, we consider a p-type NC in accordance with Bonnet *et al.*[24] since the ET occurs via the HOMO of the CsCoFe NC. Consequently, we measure the HOPG/CsCoFe diode ($\phi_{B1}$) at V>0 and the CsCoFe/PtIr tip diode ($\phi_{B2}$) at V<0 (voltage applied on HOPG, C-AFM tip grounded, see inset Fig. 5c).

The current $I_T$ between the two electrodes in the junction is written:

$$I_T = \frac{2I_{S1}I_{S2}\sinh\left(\frac{qU}{2kT}\right)}{I_{S1}e^{\frac{-qU}{2kTn_1}}+I_{S2}e^{\frac{qU}{2kTn_2}}} \; ; \; U = V - RI_T \tag{1}$$

with $q$ the elementary charge (q = 1.6 x 10$^{-19}$ C), $V$ the applied bias, $k$ the Boltzmann constant (k = 1.38 x 10$^{-23}$ J.K$^{-1}$), $T$ the temperature (here T = 293 K), $n_{1,2}$ the ideality factors and $I_{S1,S2}$ the reverse saturation currents, R the intrinsic NC resistance. The saturation current are expressed as:

$$I_{S1,S2} = S_{1,2}A^*T^2 \exp\left(-\frac{\Phi_{B1,B2}}{kT}\right) \tag{2}$$

with $S_{1,2}$ the areas of the junctions, $A^*$ the Richardson constant ($A^*$ = 1.2 x 10$^6$ A.m$^{-2}$.K$^2$ for free carriers[47]) and $\Phi_{B1,B2}$ the effective SBs. $S_1$ corresponding to the surface area of the NC with the HOPG substrate was fixed to the nominal NC base surface value (square of the NC nominal size). $S_2$ corresponds to the contact surface of the NC with the C-AFM tip, we estimate a surface $S_2$ ~ 16 nm² for a loading force of 15 nN using a Hertzian model[48] (see details in the Supporting Information Section 4).

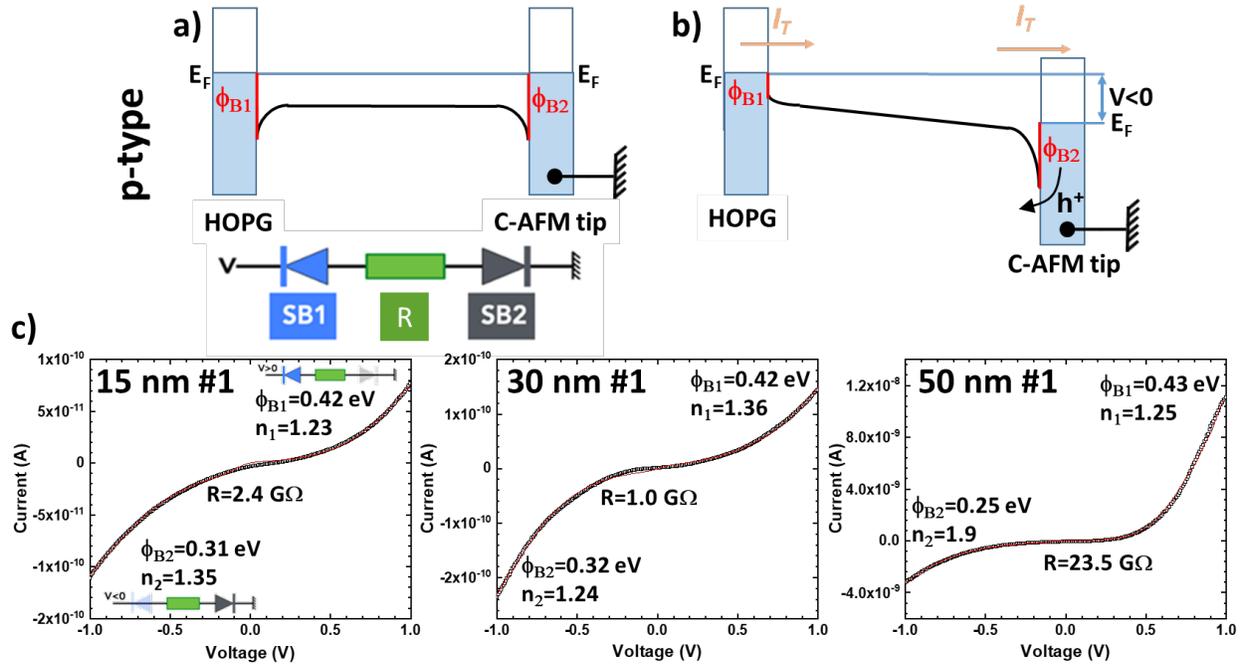

*Figure 5.* (a) Energy diagram of the HOPG/NC/C-AFM tip junction with two SBs identified with red lines without applied bias for a p-type material, and the equivalent electrical circuit with the two Schottky diodes in opposite direction and the intrinsic resistance of the NC. (b) Energy diagram of the same junction at V<0 applied on the HOPG substrate. (c) Average $\bar{I}(V)$ curves (black line) corresponding to the C-AFM 2D current-voltage histograms presented in Fig. 3 for the three nominal sizes of NC. Fits using a double SB model (Eqs 1 and 2) are shown as red lines. In the inset, the shaded diodes are in the forward regime. At V>0 (V<0, respectively), the measured current is the reverse current of the HOPG/CsCoFe (CsCoFe/PtIr tip, respectively) diodes, respectively. The fits converged with a coefficient of determination ($R^2$) values between 0.98-099.

The mean $\bar{I}(V)$ traces were obtained for the different NCs starting from the C-AFM 2D current-voltage histograms presented in Figure 3, and are shown in Figure 5. These mean $\bar{I}(V)$ traces were fitted with the double SB model (Eqs. 1 and 2) with the adjusting parameters: the effective SBs ($\Phi_{B1,B2}$), the ideality factors ($n_{1,2}$) and the intrinsic resistance R. The mean $\bar{I}(V)$ curves are well fitted (Figure 4c) by this model. A random sample of 20

I(V)s curves in the complete datasets (Fig. 3 and Figs. S3 in the Supporting Information) were also fitted by this model to obtain an estimation of the statistical distribution of the parameters. These statistical analyses (summarized in Table 2) confirm the results simply used by fitting the mean Ī(V).

| NC size (nm) | 15 | | 30 | | | 50 | | |
|---|---|---|---|---|---|---|---|---|
| | #1 | #2 | #1 | #2 | #3 | #1 | #2 | #3 |
| $\Phi_{B1}$ (eV) | 0.43±0.02 | 0.39±0.02 | 0.43±0.02 | 0.40±0.02 | 0.37±0.02 | 0.43±0.02 | 0.41±0.02 | 0.40±0.02 |
| $\Phi_{B2}$ (eV) | 0.31±0.02 | 0.28±0.02 | 0.31±0.02 | 0.25±0.02 | 0.24±0.02 | 0.25±0.02 | 0.23±0.02 | 0.26±0.02 |
| $n_1$ | 1.21±0.04 | 1.23±0.04 | 1.37±0.07 | 1.19±0.09 | 1.17±0.07 | 1.25±0.04 | 1.22±0.09 | 1.29±0.05 |
| $n_2$ | 1.33±0.05 | 1.50±0.05 | 1.25±0.02 | 1.64±0.11 | 1.23±0.09 | 1.86±0.14 | 1.43±0.07 | 1.24±0.02 |
| R (Ω) | 1.75±0.72 x$10^9$ | 2.0±1.3 x$10^7$ | 8.6±3.3 x$10^8$ | 3.7±1.8 x$10^7$ | 4.1±1.9 x$10^7$ | 1.6±0.7 x$10^7$ | 1.7±1.1 x$10^7$ | 1.5±0.3 x$10^8$ |

**Table 2**. *Fitted values of the SB heights at the HOPG/NC ($\Phi_{B1}$) and NC/tip ($\Phi_{B2}$) interfaces, the ideality factors ($n_1$ and $n_2$) and intrinsic NC resistance (R) for all the datasets. The table summarizes the mean values with the statistical dispersion, standard deviation (detailed in Supporting Information Section 5).*

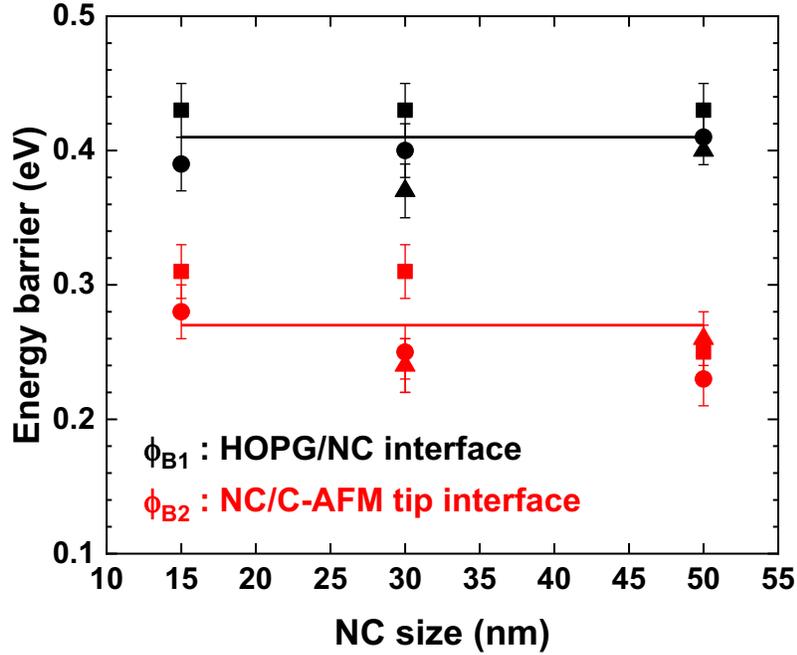

***Figure 6.*** *Barrier heights $\Phi_{B1}$ (HOPG/NC) and $\Phi_{B2}$ (NC/C-AFM tip) extracted from the fits on all the I(V) dataset: dataset #1 shown in Fig. 3 (square symbols), dataset #2 in Fig. S3 (circle symbols), dataset #3 in Fig. S3 (triangle symbols), see the Supporting Information, Section 3, for the corresponding 2D histograms and mean $\bar{I}(V)$ fits. Lines correspond to the mean values for $\Phi_{B1}$ and $\Phi_{B2}$.*

Figure 6 shows the SBH extracted from all the I(V) datasets. At the HOPG/NC interface, the energy barrier height is constant with a mean value $\Phi_{B1}$ = 0.41 ± 0.02 eV. At NC/C-AFM tip interface, we measured more disperse values in function of the NC size. This feature can be rationalized because the C-AFM tip contact is flawed by various uncontrolled fluctuations. For example, the exact shape of the surface/tip contact (and thus its exact area value) depends on the precise shape of the tip (which can evolve during the successive C-AFM measurements), it also depends on the stiffness of the NC (not precisely known for CsCoFe NCs), but for Ni/[Fe(CN)$_6$] it was observed that small NCs are stiffer than larger ones,[49] see details in section 4 in the Supporting Information). Also, albeit we used the same nominal loading force for all the measurements, the exact force applied by the C-AFM can vary and the pressure (around 1 GPa) at the tip/sample surface can also vary. It is known that the structural and electronic structures of PBA (and likely their ET properties) are dependent on the pressure.[50] Another origin of the larger dispersion of the SBH at the NC/C-AFM tip interface could also come from the presence of the water meniscus (since measurements were done in air). It is known that the presence of water or humidity modifies the electronic properties of devices based on molecular nano-objects.[51,52,53] Albeit the measurements were done in an air-conditioned laboratory (see the Experimental Section), we cannot completely exclude that the fluctuations of the water meniscus (quantity of water and its ionic impurity contents) induce additional dispersion of the SBH values.

Thus, we conclude that this larger dispersion of the $\Phi_{B2}$ values are related to a less controlled interface at the C-AFM tip and that the SBH at the NC/C-AFM tip interface is also independent on the NC size with a mean value $\Phi_{B2}$ = 0.27 ± 0.03 eV. We also notice that the barrier height $\Phi_{B1}$ is always slightly higher than the barrier height $\Phi_{B2}$ by about 0.14 eV. This difference can be ascribed to the lower work function for the HOPG electrode (W = 4.47 eV)[54] than for the PtIr electrode (W = 4.86 eV).[55] The estimated SBH values are in agreement with previous works. From first-principles relativistic many-electron calculations of Fe, Co, and Ni ferrocyanide nanocrystals, the HOMO of CsCoFe is associated to the $Fe^{2+}$-$t_{eg}$ orbital at 0.2 - 0.3 eV below the Fermi energy level.[39] From the I-V measurements on stacked layers of 15 nm CsCoFe NCs probed by C-AFM, Bonnet *et al.*[24] have shown that the ET in these samples involved the HOMO, which was measured at 0.42 - 0.55 eV with respect to the Fermi energy of the electrodes. Concerning the ideality factors, we do not observe significant trends with the size of the NC (Figure 5, Figs S3 and S4 in the Supporting Information). From the analysis of all the datasets, we deduce $n_1$ = 1.23 ± 0.07 and $n_2$ = 1.47 ± 0.27 (Fig. S4). The $n_2$ values are slightly higher than the $n_1$ values, indicating that the HOPG/C-AFM interface is a less ideal than at the HOPG, in agreement with the discussion about the fluctuations of the SBH values (*vide supra*). In both cases, the n value deviates from ideality (n=1) which is due to any inhomogeneity, impurities and defects at the interface. For instance, at the HPOG/NC interface, the inhomogeneity of the electrical properties of the HOPG surface itself at different sheets, ribbons, step-edges[56] can play a role. At the NC/C-AFM interface the presence of impurities on the tip surface, water meniscus at the tip/surface interface can also play a role as mentioned above.

We note that the energetics of the PBA NC devices, deduced by the two models, are in reasonable agreement, with almost the same values for the SBH (position of the VB below the electrode Fermi energy) and the $\varepsilon_0$ values (position of the HOMO below the electrode Fermi energy). This is consistent with the fact that the NCs are likely fully depleted (no band bending in the NCs). The very small C-AFM contact area (about 16 nm$^2$, see the Supporting Information) makes capacitance measurements impossible (capacitance $< \sim 10^{-18}$ F) in such C-AFM/NC/HOPG devices to check whether or not the capacitance is constant with the applied voltage (as expected for the fully depleted case). However, as evidenced by combining DSB model analysis and capacitance measurements in macroscopic Schottky diodes with few tens of nanometers thick semiconducting layers,[57,58] we can reasonably assume that this is also the case for the PBA single NCs. This feature makes the comparison of the interface energy barriers between the SEL and DSB models more pertinent in the absence of band bending in the Schottky diodes.

Figure 7a summarizes the estimated NC intrinsic resistance R for the whole datasets. From these values, we also estimate the intrinsic NC conductivity (Fig. 7b) using a simple equation and an idealized truncated pyramidal geometry for the HOPG/NC/PtIr tip with strongly asymmetric contact areas (see section 7 in the Supporting Information).

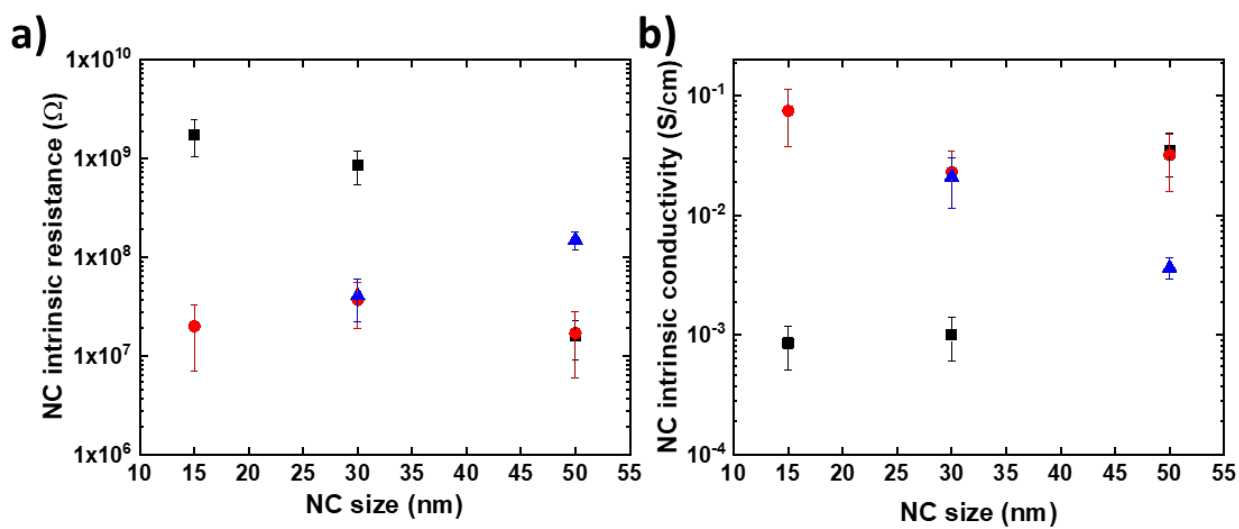

***Figure 7.*** *Evolution of the resistance and conductivity of the CsCoFe NCs extracted from the fits on the I(V) dataset: dataset #1 shown in Fig. 3 (black square symbols), dataset #2 in Fig. S3 (red circle symbols) and dataset #3 in Fig. S3 (blue triangle symbols).*

The main feature is that the intrinsic electron conductivity is dispersed with values between ~ $5 \times 10^{-2}$ – $5 \times 10^{-4}$ S/cm for the 15 nm NCs. A less dispersed conductivity between $2 \times 10^{-2}$ – $4 \times 10^{-3}$ S/cm is estimated for the 50 nm NCs. However, there is no clear correlation between the NC intrinsic resistance, nor the conductivity, with the NC size (i.e. no clear increase or decrease with size). This dispersion observed for the conductivity values on some nanocrystals mainly reflects the intrinsic dispersion of the NC structure probably inherent to the local inhomogeneities of the structure and/or chemical composition of these nano-objects. We also note that the use of an idealized truncated geometry with the nominal NC size, instead of the real shape and size, is likely to induce some dispersion of the calculated conductivity. This dispersion of the structure is inherent to the synthesis route used to fabricate these nano-objects. A full statistical, much more time-consuming, study on a larger number of nanocrystals, would be considered in the future in order to better understand this conductivity dispersion.

Nevertheless, compared to sparse data in literature, the conductivity values reported above on individual nanocrystals are higher by up to 5 decades than those measured on PBA films: $1.5 \times 10^{-8}$ to $8 \times 10^{-7}$ S/cm for NaCo[Fe(CN$_6$] films[42] and $5 \times 10^{-6}$ to $6 \times 10^{-3}$ S/cm for various Fe[Fe(CN$_6$], Fe [Ru(CN$_6$], KRu[Fe(CN$_6$] films[43]. This work shows the importance of considering the nanocrystal individually to obtain a better estimation of its electronic properties and in particular a better estimation of its intrinsic electronic conductivity.

**Conclusion**

We have conducted a detailed analysis of the electronic transport properties of CsCoFe nanocrystals with nominal size between 15 and 50 nm. We have estimated the Schottky barrier heights at the interface with HOPG and PtIr tip of a conductive AFM. We show that they are independent of the nanocrystal size with a value $0.41 \pm 0.02$ eV at the HOPG contact and $0.27 \pm 0.03$ eV at PtIr C-AFM tip contact. These values of the hole injection barrier heights (PBA valence band offset with respect to the electrode Fermi energy) are consistent with a quantum transport model that estimates hole barrier heights of ~ 0.3 - 0.5 eV. The intrinsic electron conductivity of these CsCoFe nanocrystals was also estimated, with dispersed values between ~ $5 \times 10^{-4}$ S/cm and $2 \times 10^{-2}$ S/cm without any correlation with the NC shape. The origins of the conductivity inhomogeneity remain to be established, but they might be related to the dispersion of the structure and/or chemical composition of the characterized nanocrystals.

**Author contributions**

H.T. prepared the samples (NC deposition on HOPG) and carried out all the C-AFM and TM-AFM measurements as part of his PhD thesis (supervised by S.L. and T.M.). L.C., S.M. and T.M. synthesized the PBA NCs. H.T., S.L. and D.V. analyzed the data and did the modelization. S.L. wrote the manuscript with the contributions from all authors. All authors have given approval of the final version of the manuscript.

**Conflicts of interest**

There are no conflicts to declare.

**Acknowledgements**

The IEMN SPM characterization facilities are partly supported by the French Network Renatech. This work has been financially supported by the French National Research Agency (ANR), project SPINFUN ANR-17-CE24-0004. We acknowledge D. Deresmes for his valuable help with the C-AFM instrument, and Anne Duchêne and Clément Jamme for the TOC graphic.
**Conflicts of interest**

There are no conflicts to declare.

**Acknowledgements**

The IEMN SPM characterization facilities are partly supported by the French Network Renatech. This work has been financially supported by the French National Research Agency (ANR), project SPINFUN ANR-17-CE24-0004. We acknowledge D. Deresmes for his valuable help with the C-AFM instrument, and Anne Duchêne and Clément Jamme for the TOC graphic.

**References**

1.	Catala L, Mallah T. Nanoparticles of Prussian blue analogs and related coordination polymers: From information storage to biomedical applications. *Coord Chem Rev*. Sep 2017;346:32-61. doi:10.1016/j.ccr.2017.04.005
2.	Huang YL, Ren SQ. Multifunctional Prussian blue analogue magnets: Emerging opportunities. *Applied Materials Today*. Mar 2021;22100886. doi:10.1016/j.apmt.2020.100886
3.	Reguera L, Krap CP, Balmaseda J, Reguera E. Hydrogen storage in copper Prussian blue analogues: Evidence of H-2 coordination to the copper atom. *J Phys Chem C*. Oct 2008;112(40):15893-15899. doi:10.1021/jp803714j
4.	Kaye SS, Long JR. Hydrogen storage in the dehydrated Prussian blue analogues M-3 Co(CN)(6) (2) (M = Mn, Fe, Co, Ni, Cu, Zn). *J Am Chem Soc*. May 2005;127(18):6506-6507. doi:10.1021/ja051168t
5.	Chapman KW, Southon PD, Weeks CL, Kepert CJ. Reversible hydrogen gas uptake in nanoporous Prussian Blue analogues. *Chem Commun*. 2005;(26):3322-3324. doi:10.1039/b502850g
6.	Li WJ, Han C, Cheng G, Chou SL, Liu HK, Dou SX. Chemical Properties, Structural Properties, and Energy Storage Applications of Prussian Blue Analogues. *Small*. Aug 2019;15(32)1900470. doi:10.1002/smll.201900470
7.	Yi HC, Qin RZ, Ding SX, et al. Structure and Properties of Prussian Blue Analogues in Energy Storage and Conversion Applications. *Adv Funct Mater*. Feb 2021;31(6)2006970. doi:10.1002/adfm.202006970
8.	Qian JF, Wu C, Cao YL, et al. Prussian Blue Cathode Materials for Sodium-Ion Batteries and Other Ion Batteries. *Advanced Energy Materials*. Jun 2018;8(17)1702619. doi:10.1002/aenm.201702619
9.	Xu YX, Zheng SS, Tang HF, Guo XT, Xue HG, Pang H. Prussian blue and its derivatives as electrode materials for electrochemical energy storage. *Energy Storage Materials*. Oct 2017;9:11-30. doi:10.1016/j.ensm.2017.06.002

10. Tannous C, Comstock RL. Magnetic Information-Storage Materials. In: Kasap S, Capper P, eds. *Springer Handbook of Electronic and Photonic Materials*. Springer International Publishing; 2017:1-1.
11. Matos-Peralta Y, Antuch M. Review-Prussian Blue and Its Analogs as Appealing Materials for Electrochemical Sensing and Biosensing. *J Electrochem Soc*. Nov 2019;167(1)037510. doi:10.1149/2.0102003jes
12. Nayebi B, Niavol KP, Kim SY, et al. Prussian blue-based nanostructured materials: Catalytic applications for environmental remediation and energy conversion. *Molecular Catalysis*. Sep 2021;514111835. doi:10.1016/j.mcat.2021.111835
13. Kamachi Y, Zakaria MB, Torad NL, et al. Hydrogels Containing Prussian Blue Nanoparticles Toward Removal of Radioactive Cesium Ions. *J Nanosci Nanotechno*. Apr 2016;16(4):4200-4204. doi:10.1166/jnn.2016.12607
14. Lee KM, Kawamoto T, Minami K, et al. Improved adsorption properties of granulated copper hexacyanoferrate with multi-scale porous networks. *RSC Advan*. 2016;6(20):16234-16238. doi:10.1039/c5ra25388h
15. Busquets MA, Estelrich J. Prussian blue nanoparticles: synthesis, surface modification, and biomedical applications. *Drug Discovery Today*. Aug 2020;25(8):1431-1443. doi:10.1016/j.drudis.2020.05.014
16. Buser HJ, Schwarzenbach D, Petter W, Ludi A. Crystal-Structure of Prussian Blue - $Fe_4[Fe(CN)_6]_3 \cdot xH_2O$. *Inorg Chem*. 1977;16(11):2704-2710.
17. Shriver DF, Brown DB. Environment of interstitial ions in a Prussian blue analog, $Co_3[Co(CN)_6]_2$. *Inorg Chem*. 1969/01/01 1969;8(1):42-46. doi:10.1021/ic50071a010
18. Brown DB, Shriver DF. Structures and solid-state reactions of Prussian blue analogs containing chromium, manganese, iron, and cobalt. *Inorg Chem*. 1969/01/01 1969;8(1):37-42. doi:10.1021/ic50071a009
19. Brinzei D, Catala L, Louvain N, et al. Spontaneous stabilization and isolation of dispersible bimetallic coordination nanoparticles of CsxNi Cr(CN)(6) (y). *J Mater Chem*. 2006;16(26):2593-2599. doi:10.1039/b605221e
20. Catala L, Brinzei D, Prado Y, et al. Core-Multishell Magnetic Coordination Nanoparticles: Toward Multifunctionality on the Nanoscale. *Angew Chem Int Ed*. 2009;48(1):183-187.
21. Cobo S, Molnar G, Carcenac F, et al. Thin Films of Prussian Blue: Sequential Assembly, Patterning and Electron Transport Properties at the Nanometric Scale. *J Nanosci Nanotechno*. Aug 2010;10(8):5042-5050. doi:10.1166/jnn.2010.2430
22. Pajerowski DM, Watanabe T, Yamamoto T, Einaga Y. Electronic conductivity in Berlin green and Prussian blue. *Phys Rev B*. Apr 2011;83(15)153202. doi:10.1103/PhysRevB.83.153202
23. Lefter C, Tan R, Dugay J, et al. Light induced modulation of charge transport phenomena across the bistability region in Fe(Htrz)(2)(trz) (BF4) spin crossover micro-rods. *PCCP*. 2015;17(7):5151-5154. doi:10.1039/c4cp05250a
24. Bonnet R, Lenfant S, Mazerat S, Mallah T, Vuillaume D. Long-range electron transport in Prussian blue analog nanocrystals. *Nanoscale*. Oct 2020;12(39):20374-20385. doi:10.1039/d0nr06971j
25. Datta S. *Electronic transport in mesoscopic systems*. Cambridge university press; 1997.

26.   Cuevas JC, Scheer E. *Molecular electronics: an introduction to theory and experiment*. World Scientific; 2010.
27.   Tanaka I, Kamiya I, Sakaki H. Local surface band modulation with MBE-grown InAs quantum dots measured by atomic force microscopy with conductive tip. *Journal of crystal growth*. 1999;201:1194-1197.
28.   Oh J, Nemanich RJ. Current–voltage and imaging of TiSi 2 islands on Si (001) surfaces using conductive-tip atomic force microscopy. *Journal of applied physics*. 2002;92(6):3326-3331.
29.   Rezeq Md, Abbas Y, Wen B, Wasilewski Z, Ban D. Direct detection of electronic states for individual indium arsenide (InAs) quantum dots grown by molecular beam epitaxy. *Applied Surface Science*. 2022;590:153046.
30.   Grillo A, Di Bartolomeo A. A Current-Voltage Model for Double Schottky Barrier Devices. *Advanced Electronic Materials*. Feb 2021;7(2)2000979. doi:10.1002/aelm.202000979
31.   Rhoderick EH, Williams RH. *Metal-semiconductor contacts*. vol 129. Clarendon press Oxford; 1988.
32.   Tung RT. Electron transport of inhomogeneous Schottky barriers. *Applied Physics Letters*. 1991;58(24):2821-2823. doi:10.1063/1.104747
33.   Tung RT. Electron transport at metal-semiconductor interfaces: General theory. *Physical Review B*. 06/15/ 1992;45(23):13509-13523. doi:10.1103/PhysRevB.45.13509
34.   Ruffino F, Grimaldi MG, Giannazzo F, Roccaforte F, Raineri V. Size-dependent Schottky barrier height in self-assembled gold nanoparticles. *Applied physics letters*. 2006;89(24)
35.   Hasegawa H, Sato T, Kasai S. Unpinning of Fermi level in nanometer-sized Schottky contacts on GaAs and InP. *Applied surface science*. 2000;166(1-4):92-96.
36.   Trinh L, Zerdane S, Mazerat S, et al. Photoswitchable 11 nm CsCoFe Prussian Blue Analogue Nanocrystals with High Relaxation Temperature. *Inorg Chem*. Sep 2020;59(18):13153-13161. doi:10.1021/acs.inorgchem.0c01432
37.   Prado Y, Mazerat S, Riviere E, et al. Magnetization Reversal in (CsNiCrIII)-Cr-II(CN)(6) Coordination Nanoparticles: Unravelling Surface Anisotropy and Dipolar Interaction Effects. *Adv Funct Mater*. Sep 10 2014;24(34):5402-5411.
38.   Wojdeł JC, de P. R. Moreira I, Bromley ST, Illas F. On the prediction of the crystal and electronic structure of mixed-valence materials by periodic density functional calculations: The case of Prussian Blue. *The Journal of Chemical Physics*. 2008;128(4):044713. doi:10.1063/1.2824966
39.   Watanabe S, Sawada Y, Nakaya M, et al. Intra- and inter-atomic optical transitions of Fe, Co, and Ni ferrocyanides studied using first-principles many-electron calculations. *J Appl Phys*. Jun 2016;119(23)235102. doi:10.1063/1.4954070
40.   Rosseinsky DR, Tonge JS, Berthelot J, Cassidy JF. SITE-TRANSFER CONDUCTIVITY IN SOLID IRON HEXACYANOFERRATES BY DIELECTRIC RELAXOMETRY, VOLTAMMETRY AND SPECTROSCOPY - PRUSSIAN BLUE, CONGENERS AND MIXTURES. *Journal of the Chemical Society-Faraday Transactions I*. 1987;83:231-243. doi:10.1039/f19878300231

41. Xidis A, Neff VD. ON THE ELECTRONIC CONDUCTION IN DRY THIN-FILMS OF PRUSSIAN BLUE, PRUSSIAN YELLOW, AND EVERITT SALT. *Journal of the Electrochemical Society*. Dec 1991;138(12):3637-3642. doi:10.1149/1.2085472
42. Sato O, Kawakami T, Kimura M, Hishiya S, Kubo S, Einaga Y. Electric-field-induced conductance switching in FeCo Prussian blue analogues. *J Am Chem Soc*. Oct 2004;126(41):13176-13177. doi:10.1021/ja046329s
43. Behera JN, D'Alessandro DM, Soheilnia N, Long JR. Synthesis and Characterization of Ruthenium and Iron-Ruthenium Prussian Blue Analogues. *Chem Mater*. May 2009;21(9):1922-1926. doi:10.1021/cm900230p
44. Amirav L, Wachtler M. Nano Schottky? *Nano Letters*. Dec 2022;22(24):9783-9785. doi:10.1021/acs.nanolett.2c04150
45. Angizi S, Selvaganapathy PR, Kruse P. Graphene-silicon Schottky devices for operation in aqueous environments: Device performance and sensing application. *Carbon*. Jul 2022;194:140-153. doi:10.1016/j.carbon.2022.03.052
46. Liu H-Y, Yin J, Gao X, et al. Scalable Submicron Channel Fabrication by Suspended Nanofiber Lithography for Short-Channel Field-Effect Transistors. *Adv Funct Mater*. Feb 2022;32(6)2109254. doi:10.1002/adfm.202109254
47. Sze SM. Physics of Semiconductor Devices. Mar 9 1981:1-442.
48. Wold DJ, Frisbie CD. Fabrication and characterization of metal-molecule-metal junctions by conducting probe atomic force microscopy. *J Am Chem Soc*. 2001;123:5549-5556.
49. Felix G, Mikolasek M, Shepherd HJ, et al. Elasticity of Prussian-Blue-Analogue Nanoparticles. *European Journal of Inorganic Chemistry*. Jan 2018;(3-4):443-448. doi:10.1002/ejic.201700796
50. Bleuzen A, Cafun JD, Bachschmidt A, et al. CoFe Prussian Blue Analogues under Variable Pressure. Evidence of Departure from Cubic Symmetry: X-ray Diffraction and Absorption Study. *J Phys Chem C*. Nov 2008;112(45):17709-17715. doi:10.1021/jp805852n
51. Long DP, Lazorcik JL, Mantooth BA, et al. Effects of hydration on molecular junction transport. *Nature Materials*. 2006;5:901-908.
52. Zhang X, Mcgill SA, Xiong P. Origin of the humidity sensitivity of Al/AlO(x)/MHA/Au molecular tunnel junctions. *J Am Chem Soc*. Nov 21 2007;129(46):14470-4. doi:10.1021/ja0758988
53. Smaali K, Clément N, Patriarche G, Vuillaume D. Conductance Statistics from a Large Array of Sub-10 nm Molecular Junctions. *ACS Nano*. 2012/06/26 2012;6(6):4639-4647. doi:10.1021/nn301850g
54. Hansen WN, Hansen GJ. Standard reference surfaces for work function measurements in air. *Surface Science*. Jun 2001;481(1-3):172-184. doi:10.1016/s0039-6028(01)01036-6
55. Lee NJ, Yoo JW, Choi YJ, et al. The interlayer screening effect of graphene sheets investigated by Kelvin probe force microscopy. *Appl Phys Lett*. Nov 2009;95(22)222107. doi:10.1063/1.3269597
56. Banerjee S, Sardar M, Gayathri N, Tyagi AK, Raj B. Conductivity landscape of highly oriented pyrolytic graphite surfaces containing ribbons and edges. *Phys Rev B*. Aug 2005;72(7)075418. doi:10.1103/PhysRevB.72.075418

57. Oswald J, Beretta D, Stiefel M, et al. Field and Thermal Emission Limited Charge Injection in Au–$C_{60}$–Graphene van der Waals Vertical Heterostructures for Organic Electronics. *ACS Appl Nano Mater*. 2023;6:9444-9452. doi:10.1021/acsanm.3c01090
58. Kim CH, Yaghmazadeh O, Tondelier D, Jeong YB, Bonnassieux Y, Horowitz G. Capacitive behavior of pentacene-based diodes: Quasistatic dielectric constant and dielectric strength. *Journal of Applied Physics*. 2011;109(8):083710. doi:10.1063/1.3574661

# Electronic Properties of Single Prussian Blue Analog Nanocrystals Determined by Conductive-AFM.

Hugo Therssen[1], Laure Catala[2], Sandra Mazérat[2], Talal Mallah[2], Dominique Vuillaume[1], Thierry Mélin[1], Stéphane Lenfant[1]

1. Institut d'Electronique de Microélectronique et de Nanotechnologie (IEMN), CNRS, Univ. Lille, 59652 Villeneuve d'Ascq, France.
2. Institut de Chimie Moléculaire et des Matériaux d'Orsay (ICMMO), CNRS, Université Paris-Saclay, 91400 Orsay Cedex, France.

**SUPPORTING INFORMATION**

1. Synthesis of the PBA nanocrystals
2. Current-voltage 2D histogram measured on the bare HOPG substrates
3. Supplementary I(V) datasets and analysis
4. Estimation of the contact area at the NC/C-AFM interface using a Hertzian model
5. Statistical analysis of the dataset in Fig. 3
6. Ideality factors.
7. Estimation of the NC conductivity
8. I(V) analysis using the single-energy level (SEL) model

**1. Synthesis of the PBA nanocrystals**

<u>15 nm nanocrystals (CsCoFe_15)</u>. An aqueous solution (100 ml) containing 69.4 mg (4x10$^{-3}$ ML$^{-1}$) of CsCl and 47.7 mg (2x10$^{-3}$ ML$^{-1}$) of [Co(H$_2$O)$_6$]Cl$_2$ is prepared. Then another aqueous solution (100 ml) containing 67.6 mg (2x10$^{-3}$ ML$^{-1}$) of K$_3$[Fe(CN)$_6$] is prepared. The two solutions were mixed and vigorously stirred leading to the CsCoFe_15 nanocrystals.

<u>30 nm nanocrystals (CsCoFe_30)</u>. An aqueous solution (240 ml) containing 122.2 mg (3x10$^{-3}$ ML$^{-1}$) of CsCl and 87.2 mg (1.5x10$^{-3}$ ML$^{-1}$) of [Co(H$_2$O)$_6$]Cl$_2$ is prepared. Then another aqueous solution (240 ml) containing 87.2 mg (1.5x10$^{-3}$ ML$^{-1}$) of K$_3$[Fe(CN)$_6$] is prepared.

The two solutions were added simultaneously using a peristaltic pump (1 ml/minute) onto 50 ml of the previous solution containing the 15 nm nanocrystals, completed with 20 ml of water, that play the role of seeds for the growth of the objects leading to CsCoFe_30 nanocrystals.

<u>50 nm nanocrystals (CsCoFe_50)</u>. An aqueous solution (180 ml) containing 92.4 mg (3x10$^{-3}$ ML$^{-1}$) of CsCl and 65.4 mg (1.5x10$^{-3}$ ML$^{-1}$) of [Co(H$_2$O)$_6$]Cl$_2$ is prepared. Then another aqueous solution (240 ml) containing 89.4 mg (1.5x10$^{-3}$ ML$^{-1}$) of K$_3$[Fe(CN)$_6$] is prepared.

The two solutions were added simultaneously using a peristaltic pump (1 ml/minute) onto 100 ml of the previous solution containing the 30 nm nanocrystals, completed with 40 ml of water, that play the role of seeds for the growth of the 50 nm objects leading to CsCoFe_50 nanocrystals.

The size of the objects corresponds to hydrodynamic diameter measured by DLS as depicted in the Figure S1.

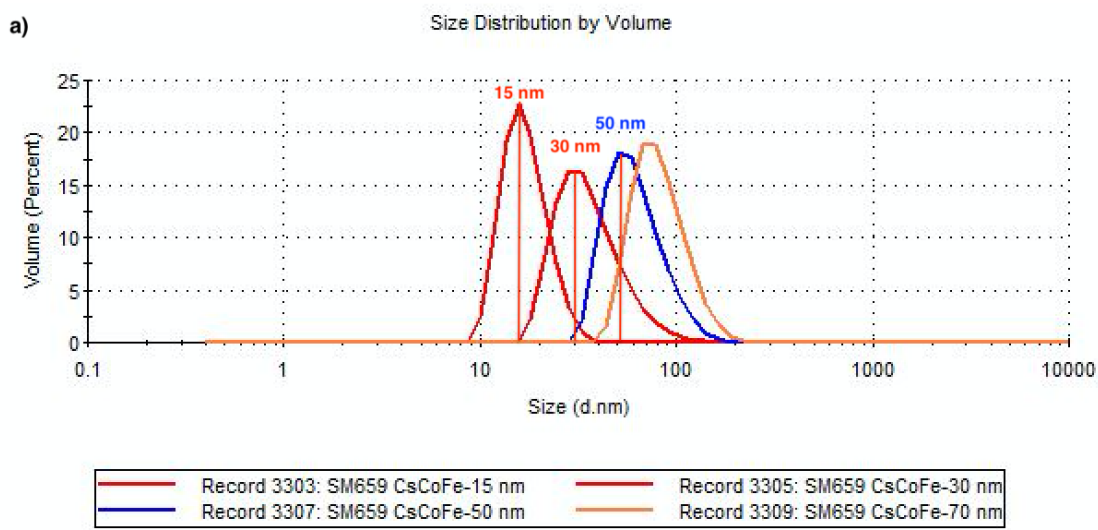

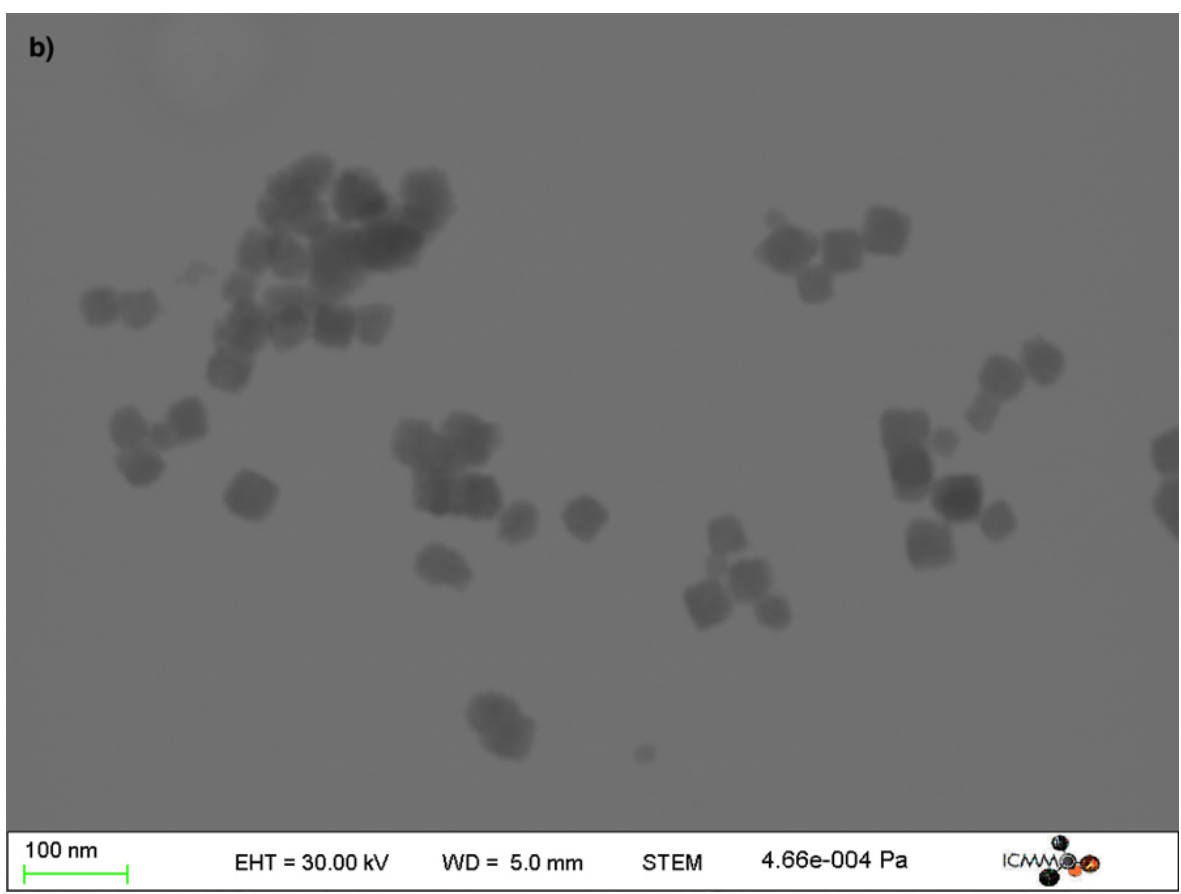

***Figure S1.*** *a) Hydrodynamic diameter of the objects for the three synthesis steps (the plot for the higher diameter objects is shown to demonstrate that this synthesis protocol can be used to prepare objects of a size larger than 50 nm, here 70 nm), b) Transmission electron*

*microscopy image of a sample of CsCoFe_50 showing the cubic shape of the nanocrystals as expected[1] and showing a relatively large distribution of size that is very narrow for the pristine objects. The large distribution is due to the two-step preparation process that leads from the 15 to the 50 nm samples.*

**2. Current-voltage 2D histogram measured on the bare HOPG substrates**

It is known that bare HOPG substrates have a large dispersion of conductance (Fig. S2) depending on the exact sheets, ribbons, step edges contacted by the tip of the C-AFM with current from tens of nA (at low voltage <1V) up to µA and larger.[2]

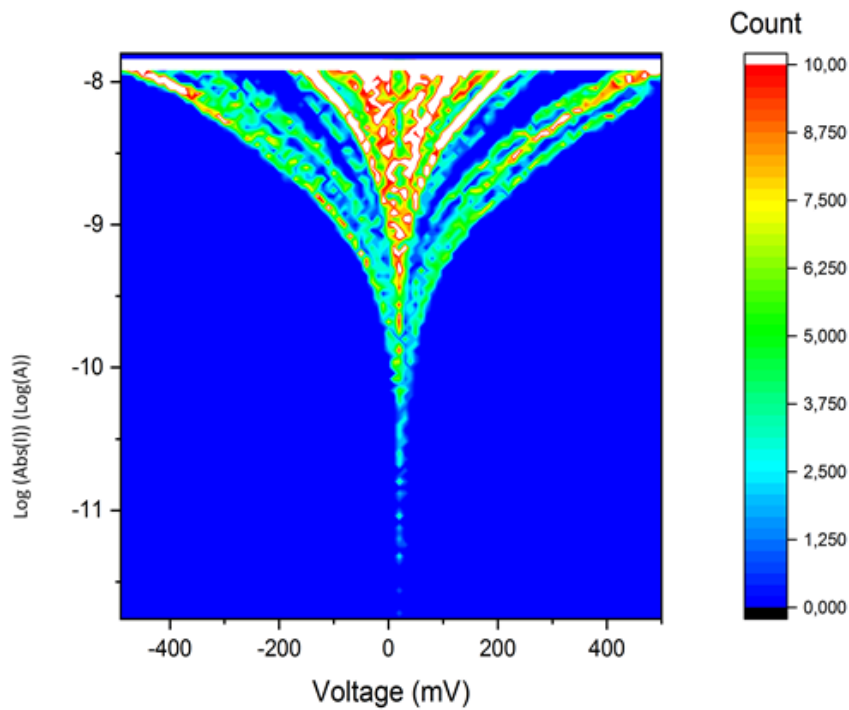

*Figure S2. 2D histogram of the HOPG substrate used for the NC deposition with a mean current measured by C-AFM; log-$\bar{I}$ superior to -8.5 (corresponding to $\bar{I} > 3 \times 10^{-9}$ A) at 0.4 V.*

## 3. Supplementary I(V) datasets and analysis

On every samples, 2 to 3 NCs were measured following the same protocol as for the dataset shown and discussed in the main text (Figs.4, 6, 7).

**Dataset #2**

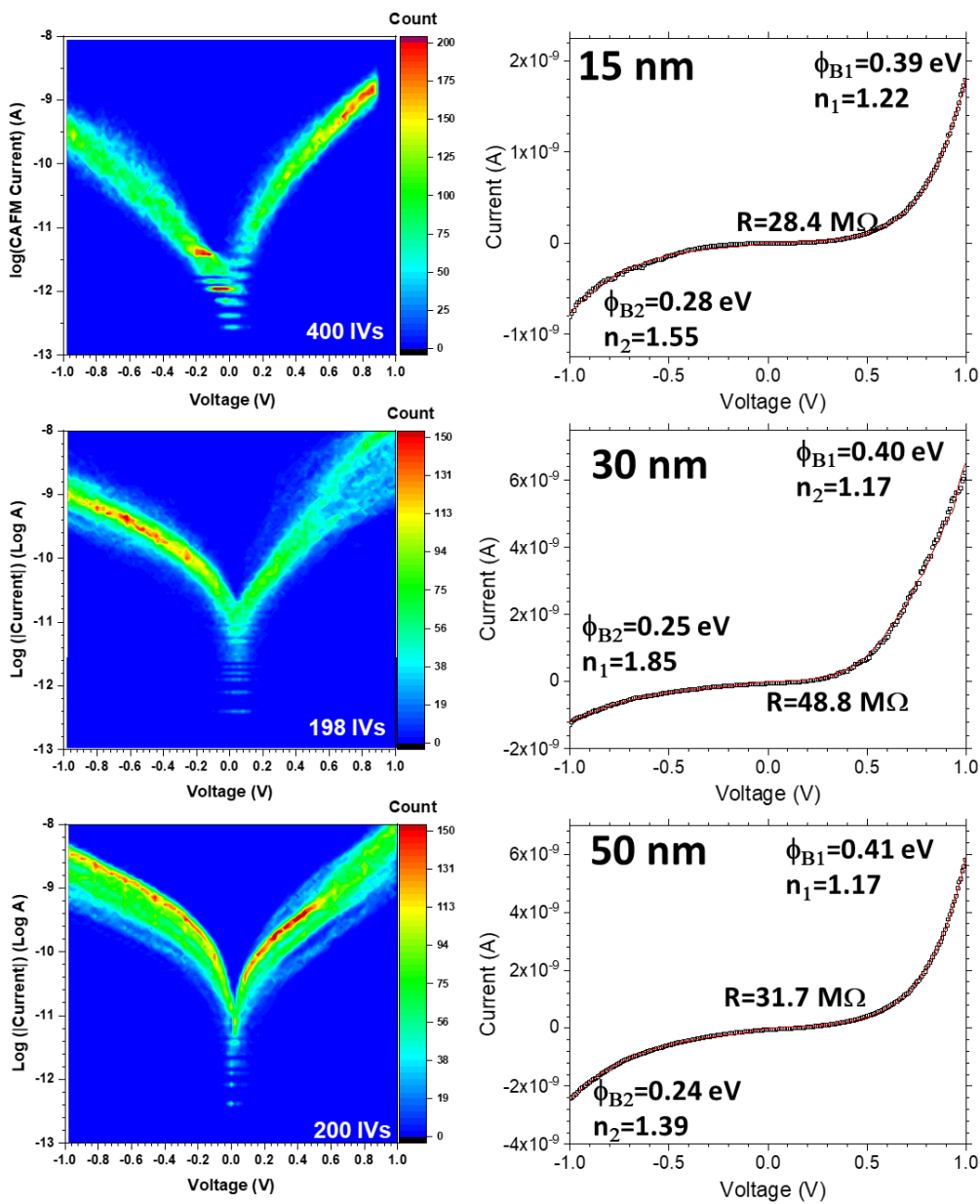

**Dataset #3**

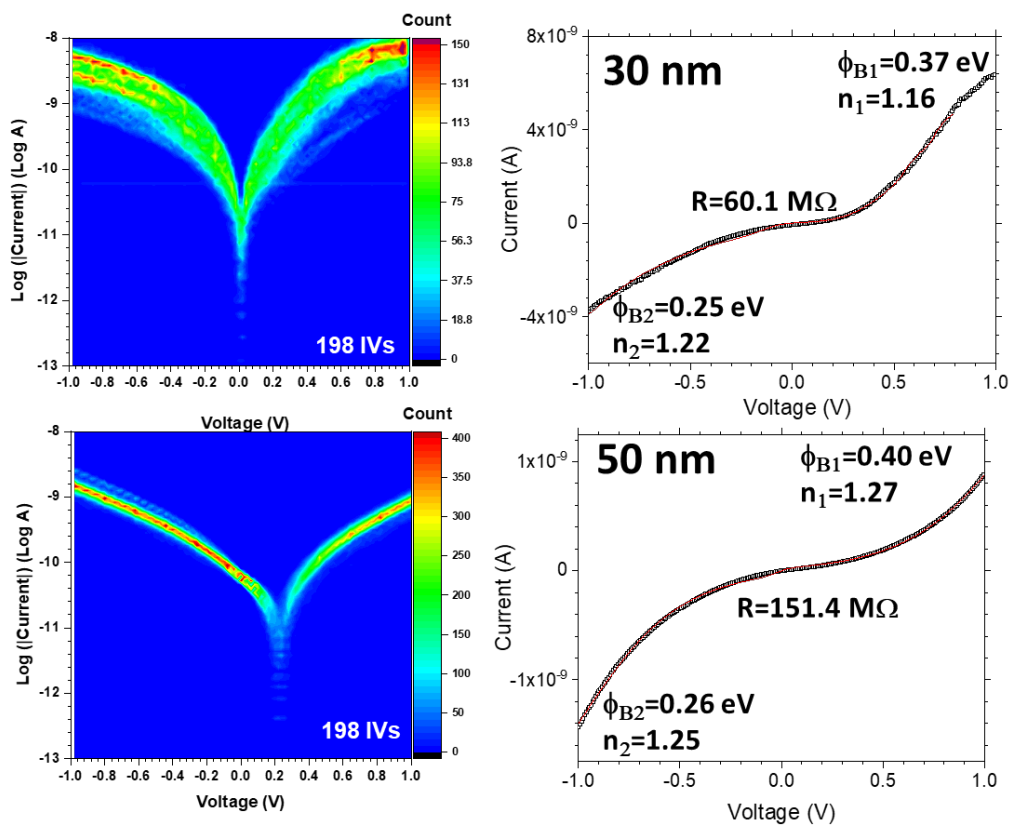

***Figure S3.*** *C-AFM 2D current-voltage I(V) histograms obtained with around 200 IVs traces for PBA NCs (Dataset #2 and #3) (Left), and the corresponding mean Ī(V) curves (black line) (Right) adjusted using the double SB model and by adjusting parameters: the effective SBs ($\Phi_{B1,B2}$), the ideality factors ($n_{1,2}$) and the intrinsic resistance R (red lines).*

## 4. Estimation of the contact area at the NC/C-AFM interface using a Hertzian model

We estimate the electrical contact surface for a loading force of 15 nN using a Hertzian model:[3]

$$S_2 = \pi \left(\frac{3R\, F_{applied}}{4E^*}\right)^{\frac{2}{3}} \quad (S1)$$

where $R$ is the radius of the C-AFM tip (fixed at 25 nm according to the manufacturer), $F_{applied}$ the tip load force (15nN), $E^*$ the reduced effective Young modulus defined as:

$$E^* = \left(\frac{1-\gamma_{NC}^2}{E_{NC}} + \frac{1-\gamma_{tip}^2}{E_{tip}}\right)^{-1} \quad (S2)$$

with $E_{NC/tip}$ and $\gamma_{NC/tip}$ the Young modulus and the Poisson ratio of the NC and the C-AFM tip respectively. The Young modulus of CsCoFe NCs is not known. A value of 43 GPa was reported for bulk material.[4] Values of about 30 GPa and 24 GPa were measured for 3 nm and 115 nm NCs of another material (Ni/[Fe(CN)$_6$]).[5]

By using the following values $E_{tip}$ = 204 GPa,[6] $E_{NC}$ = 24 GPa (by default, see above),[5] $\gamma_{tip}$ = 0.37[6] and $\gamma_{NC}$ = 0.36,[5] we estimate a surface $S_2$ ~ 16 nm². This Hertzian model gives an estimation of the NC elastic deformation close here to 0.2 nm, negligible compared to the size of the NC.

## 5. Statistical analysis of the dataset in Fig. 3 and in Fig. S3

The DSB model was fitted on all the individuals I(V) curves of the dataset to construct the statistical distribution of the model parameters shown in Fig. 3, main text (Dataset #1) and in Fig S3 (Dataset #2 and #3).

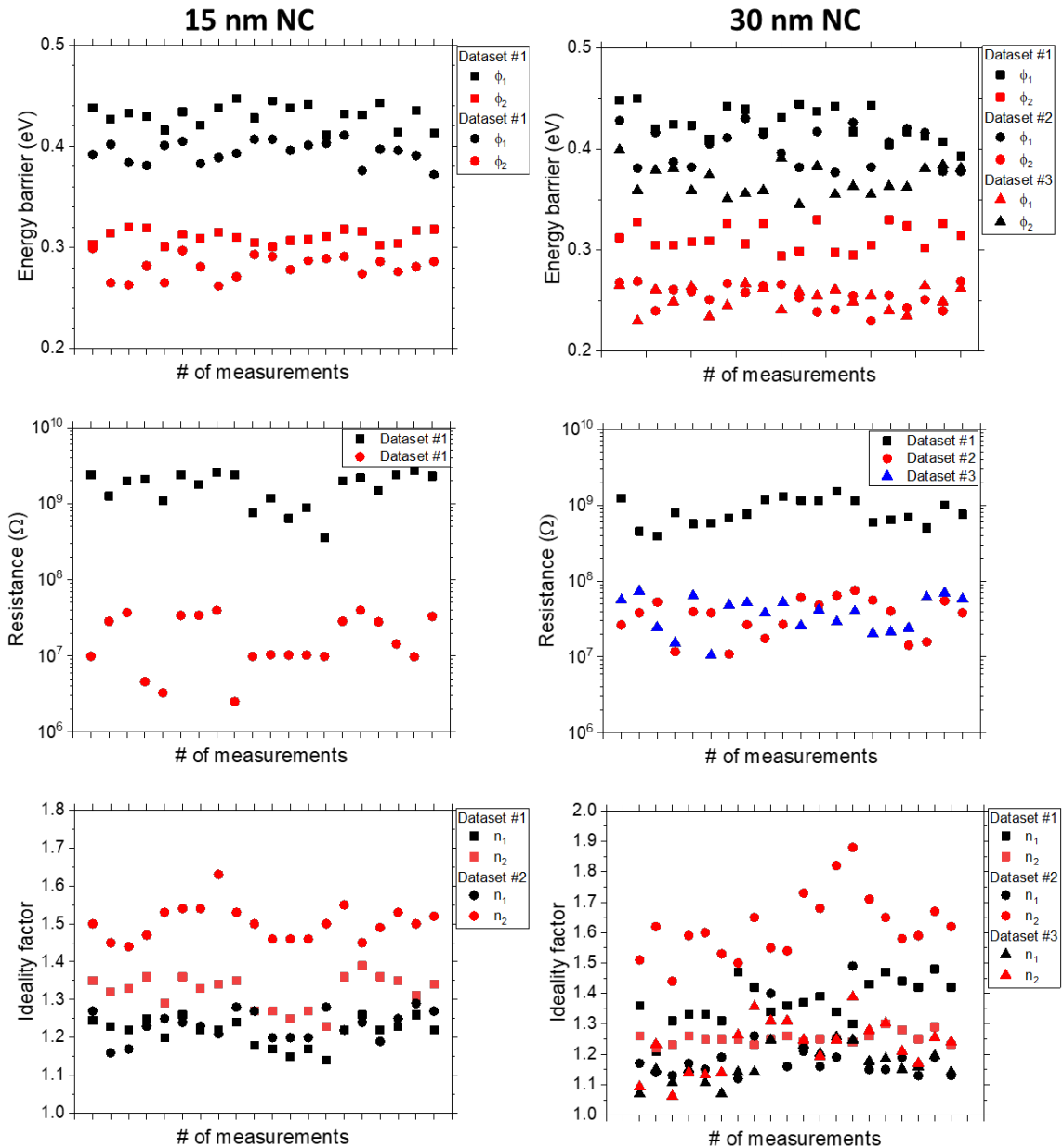

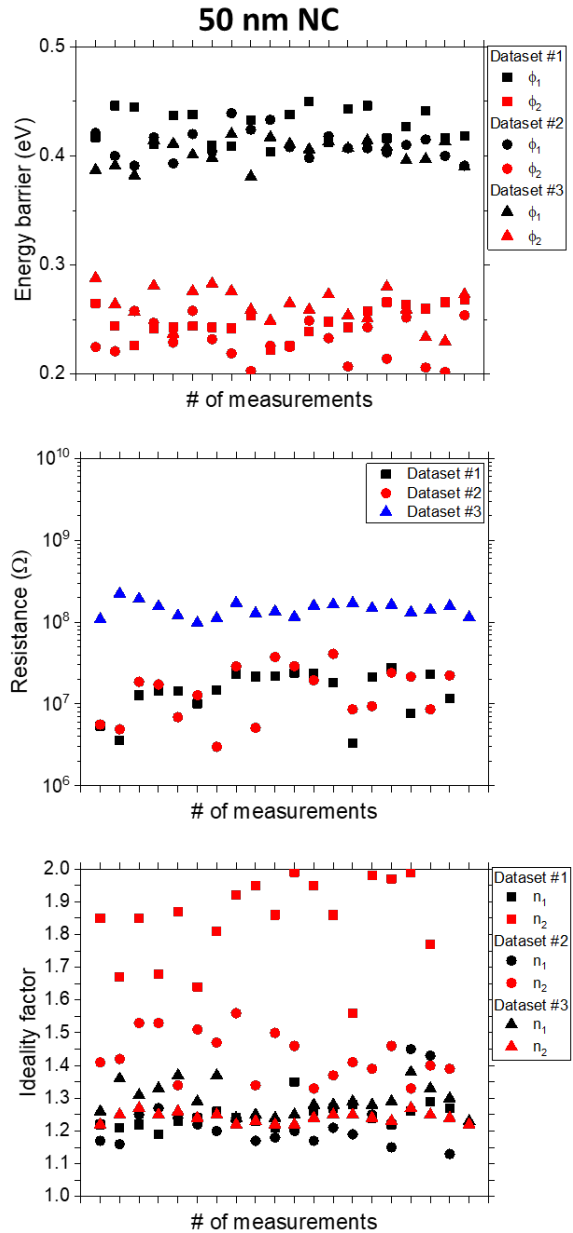

***Figure S4***. *Statistical distributions of the fitted model values obtained with the fitting with the double SB model and by adjusting parameters: the effective SBs ($\Phi_{B1,B2}$), the ideality factors ($n_{1,2}$) and the intrinsic resistance R, on a random sample of 20 I(V)s curves extract in the complete datasets presented in Fig. 3 and Fig. S3 to obtain an estimation of the statistical distribution of the parameters.*

## 6. Ideality factors.

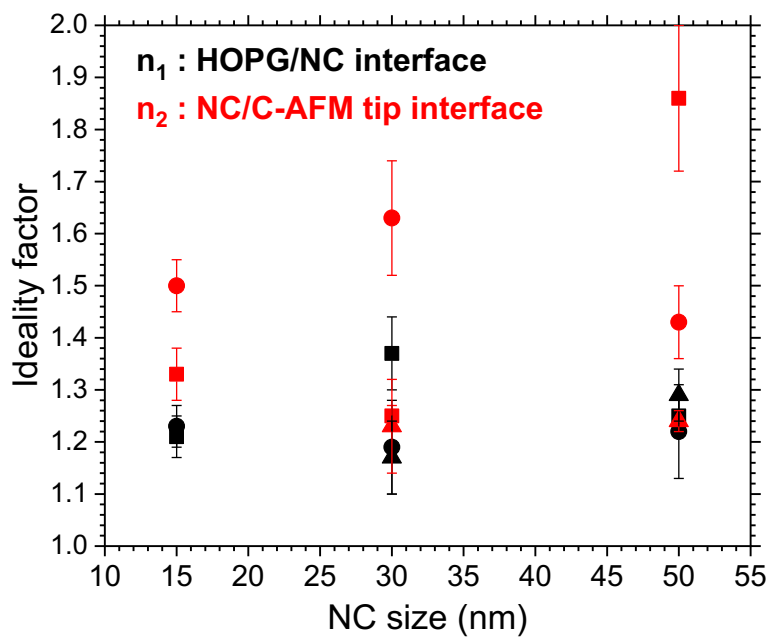

*Figure S5.* *Evolution of the ideality factor of the two CsCoFe NC Schottky diodes. The values refer to the dataset #1 shown in Fig. 3 (square symbols), dataset #2 in Fig. S3 (circle symbols), and dataset #3 in Fig. S3 (triangle symbols).*

## 7. Estimation of the NC conductivity.

The basic equation to estimate the conductivity σ from the resistance R is : $G=(R)^{-1}=\sigma\, S/L$ for a regular tube of surface S and length L, of equivalently $G=(R)^{-1}=\sigma\, V/L^2$ with V the volume of the material contacted between the two electrodes. Here, we consider a truncate pyramidal device with asymmetric contacts at the HOPG/NC and at the NC/PtIr tip (see scheme in Figure S6). For a truncated pyramid inside the nanocube of side a, the volume is given by $V = a(a^2+ar+r^2)/3$, thus we used $G=\sigma(a+r+r^2/a)/3$ to estimate the conductivity, with a the NC size, r = 4 nm.

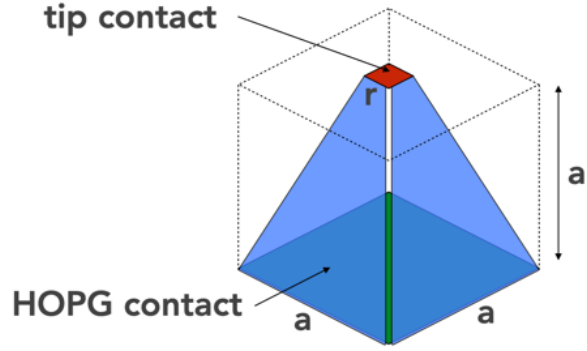

*Figure S6.* Truncate pyramidal device with asymmetric contacts at the HOPG/NC and at the NC/PtIr tip considered for the estimation of the NC conductivity.

## 8. I(V) analysis using the single-energy level (SEL) model

The single-energy level (SEL) model (Eq. 3), considers that: i) a single molecular orbital (MO) dominates the charge transport, ii) the voltage mainly drops at the molecule/electrode interface and iii) that the MO broadening is described by a Lorentzian or Breit-Wigner distribution.[7,8] The simple energy scheme (inset Fig. S7) is described by $\varepsilon_0$ the energy of the MO involved in the transport (with respect to the Fermi energy of the electrodes), $\Gamma_1$ and $\Gamma_2$ the electronic coupling energy between the MO and the electron clouds in the two electrodes, e the elementary electron charge, h the Planck constant. This analytical model reads:

$$I(V) = \frac{8e}{h}\frac{\Gamma_1\Gamma_2}{\Gamma_1+\Gamma_2}\left[\arctan\left(\frac{\varepsilon_0 + \frac{\Gamma_1}{\Gamma_1+\Gamma_2}eV}{\Gamma_1+\Gamma_2}\right) - \arctan\left(\frac{\varepsilon_0 - \frac{\Gamma_2}{\Gamma_1+\Gamma_2}eV}{\Gamma_1+\Gamma_2}\right)\right] \quad (S3)$$

This model is valid at 0 K, since the Fermi-Dirac electron distribution of the electrodes is not taken into account. However, it was shown that it can be reasonably used to fit data measured at room temperature for voltages below the transition between the off-resonant and resonant transport conditions at which the broadening of the Fermi function modify the I-V shape leading to sharpened increase of the current.[9-11] Moreover, for the sake of comparison with the I(V)s previously measured for multi-NCs devices, which were acquired between -0.5 and 0.5 V,[12] we limited the fit of Eq. S3 to this voltage window. We verified that the condition of applicability of the 0K SEL model to room temperature experimental data is satisfied (here with $\varepsilon_0 < 0.5$ eV, $\Gamma_1$ and $\Gamma_2$ around 0.1-1 meV, this condition is $|V|<0.64$ V).[13] The fits were done with the routine included in ORIGIN software (version 2019, OriginLab Corporation, Northampton, MA, USA), using the method of least squares and the Levenberg Marquardt iteration algorithm. Figure S7 shows the fits on the average I(V) of the three datasets shown Fig. 3 and Fig. S3. The fitted energy $\varepsilon_0$ and the electrode coupling energies $\Gamma_1$ and $\Gamma_2$ are summarized in Figs.S8 and S9, respectively.

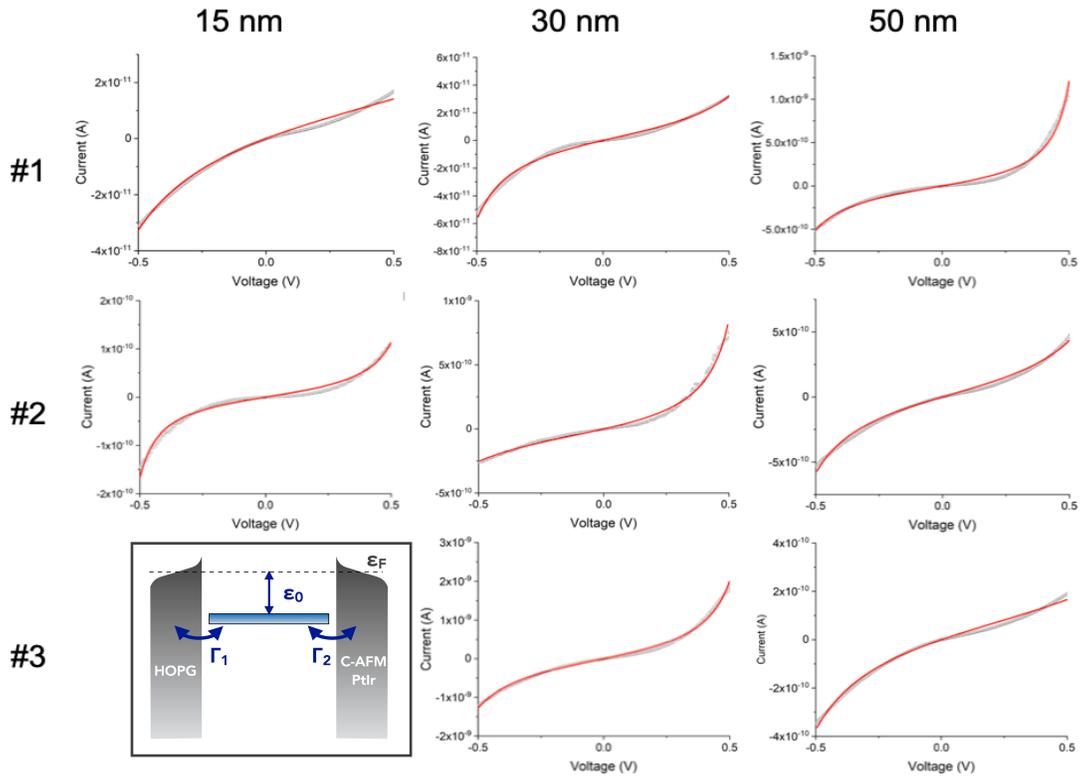

*Figure S7*. Fits of the SEL model (red lines) on the average I(V) (open squares) for the three datasets (no data at 15 nm for the dataset #3). The inset shows the energy scheme of the HOPG/NC/PtIr device with the SEL model parameters.

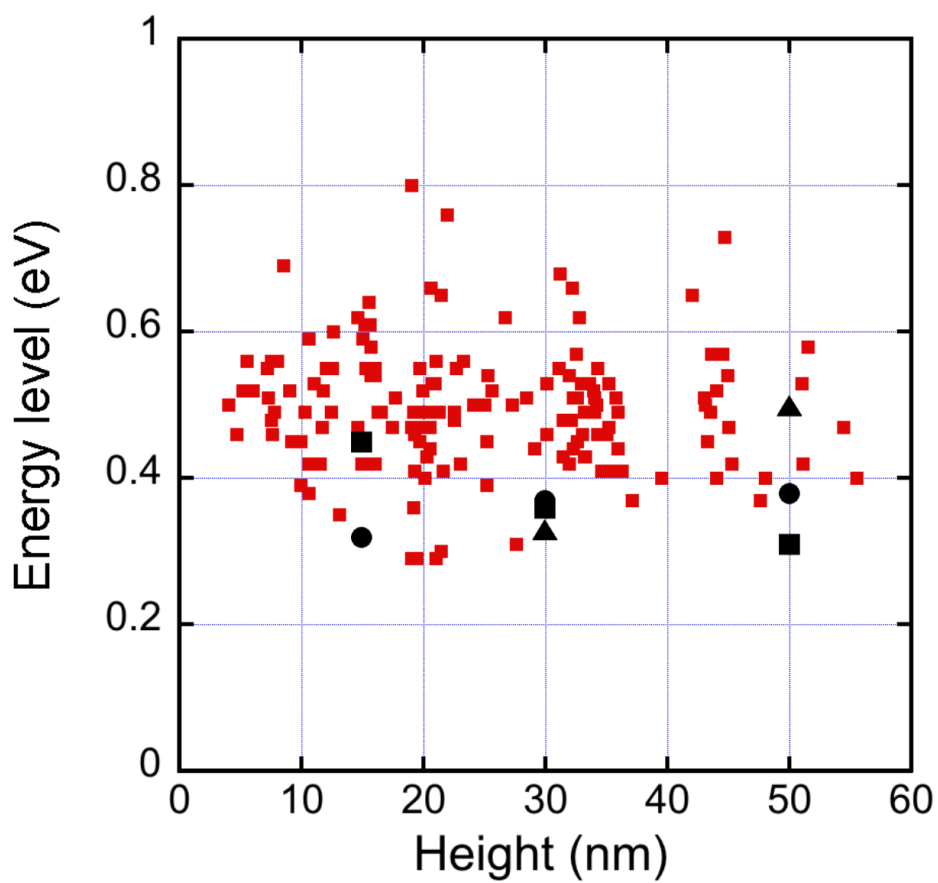

*Figure S8*. *Energy levels $\varepsilon_0$ obtained from the SEL model (fits shown in Fig. S7): dark squares (dataset #1), dark circles (dataset #2), dark triangles (dataset #3). The red symbols are the full statistics for the multi-NCs devices (from Fig. S18 in the ESI of Ref. [12]).*

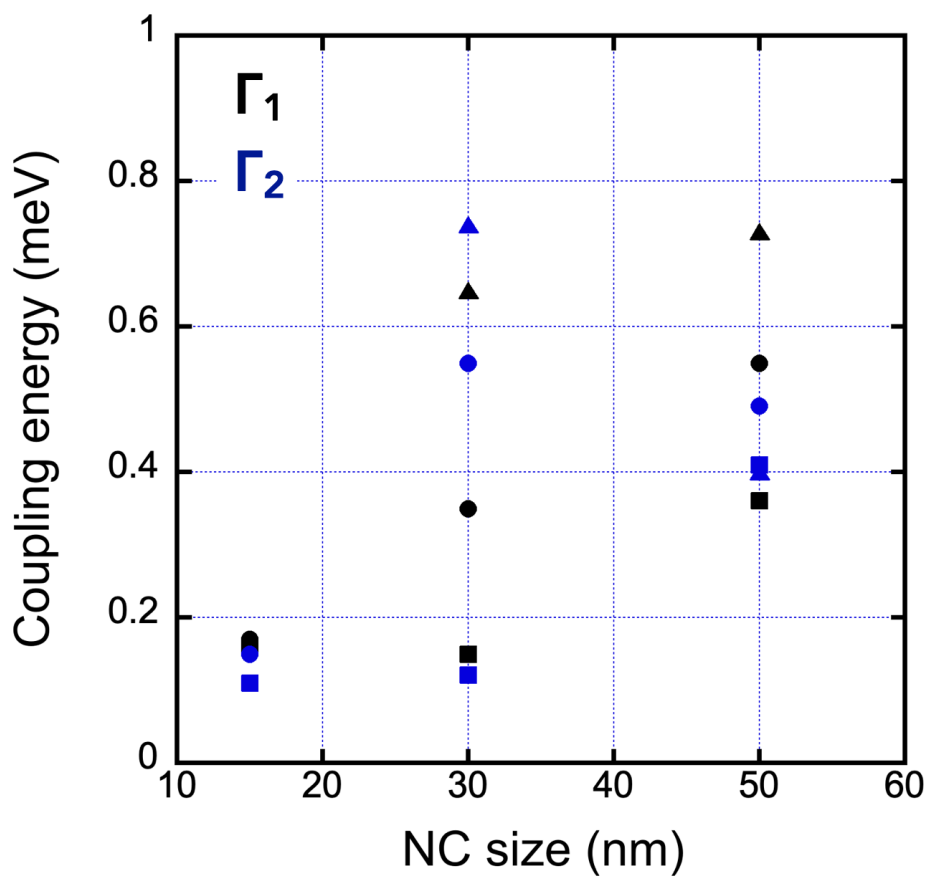

*Figure S9*. *Evolution of the electrode coupling energies Γ₁ and Γ₂ versus the NC size: square symbols (dataset #1), circle symbols (dataset #2), triangle symbols (dataset #3).*

**References**

1. L. Trinh, S. Zerdane, S. Mazerat, N. Dia, D. Dragoe, C. Herrero, E. Riviere, L. Catala, M. Cammarata, E. Collet and T. Mallah, *Inorg. Chem.*, 2020, **59**, 13153-13161.
2. S. Banerjee, M. Sardar, N. Gayathri, A. K. Tyagi and B. Raj, *Physical Review B*, 2005, **72**.
3. D. J. Wold and C. D. Frisbie, *J. Am. Chem. Soc.*, 2001, **123**, 5549-5556.
4. A. Bleuzen, J. D. Cafun, A. Bachschmidt, M. Verdaguer, P. Munsch, F. Baudelet and J. P. Itie, *Journal of Physical Chemistry C*, 2008, **112**, 17709-17715.
5. G. Felix, M. Mikolasek, H. J. Shepherd, J. Long, J. Larionova, Y. Guari, J. P. Itie, A. I. Chumakov, W. Nicolazzi, G. Molnar and A. Bousseksou, *European Journal of Inorganic Chemistry*, 2018, 443-448.
6. V. Prudkovskiy, I. Arbouch, A. Leaustic, P. Yu, C. Van Dyck, D. Guerin, S. Lenfant, T. Mallah, J. Cornil and D. Vuillaume, *Nanoscale*, 2022, **14**, 5725-5742.
7. S. Datta, *Electronic transport in mesoscopic systems*, Cambridge university press, 1997.
8. J. C. Cuevas and E. Scheer, *Molecular Electronics: An introduction to theory and experiment*, World Scientific, 2010.
9. L. Grüter, F. Cheng, T. T. Heikkilä, M. T. González, F. Diederich, C. Schönenberger and M. Calame, *Nanotechnology*, 2005, **16**, 2143-2148.
10. J. P. Bourgoin, D. Vuillaume, M. Goffman and A. Filoramo, in *Nanoscience*, eds. C. Dupas, P. Houdy and M. Lahmani, Springer, Berlin, 2007.
11. J. Brunner, Basel University, 2013.
12. R. Bonnet, S. Lenfant, S. Mazerat, T. Mallah and D. Vuillaume, *Nanoscale*, 2020, **12**, 20374-20385.
13. I. Baldea, *Phys. Chem. Chem. Phys.*, 2023, **25**, 19750-19763.